\documentclass[showpacs,preprint]{revtex4}
\usepackage{amsfonts}
\usepackage{amsmath}
\usepackage{amssymb}
\usepackage{graphicx}
\usepackage{color}
\usepackage{epsfig}
\usepackage{remreset}
\usepackage{hyperref}

\begin{document}

\title{Linear growth of entanglement entropy in holographic thermalization
captured by horizon interiors and mutual information}
\author{Yong-Zhuang Li$^{1}$}
\author{Shao-Feng Wu$^{1,2}$\footnote{%
Corresponding author. Email: sfwu@shu.edu.cn; Phone: +86-021-66136202.}}
\author{Yong-Qiang Wang$^{3}$}
\author{Guo-Hong Yang$^{1,2}$}
\pacs{11.25.Tq, 12.38.Mh, 11.25.Tq, 03.65.Ud}
\affiliation{$^{1}$Department of physics, Shanghai University, Shanghai, 200444, P. R.
China}
\affiliation{$^{2}$The Shanghai Key Lab of Astrophysics, Shanghai, 200234, P. R. China}
\affiliation{$^{3}$Institute of Theoretical Physics, Lanzhou University, Lanzhou, 730000,
P. R. China}

\begin{abstract}
We study the holographic entanglement entropy in a homogeneous falling shell
background, which is dual to the strongly coupled field theory following a
global quench. For d=2 conformal field theories, it is known that the
entropy has a linear growth regime if the scale of the entangling region
is large. In addition, the growth rate approaches a constant when the scale
increases. We demonstrate analytically that this behavior is directly
related to the part of minimal area surface probing the interior of apparent
horizons in the bulk, as well as the mutual information between two disjoint
rectangular subsystems in the boundary. Furthermore, we show numerically
that all the results are universal for the d=3 conformal field theory, the
non-relativistic scale-invariant theory and the dual theory of Gauss-Bonnet
gravity.
\end{abstract}

\maketitle

\section{Introduction}

As the most concrete realization of holographic principle, the gauge/gravity
duality \cite{Maldacena9711200} has been fruitful in revealing universal
features of strongly coupled field theories by gravitational description and
also has the\ potential ability of encoding the quantum gravity using field
theory language.

The duality has gone beyond the equilibrium and near-equilibrium processes,
relating the thermalization of far-from-equilibrium boundary gauge theories
to the gravitational collapse and the formation of black holes in the bulk.
The non-equilibrium holography is well motivated by the demand of describing
the fast thermalization of the quark gluon plasma produced in heavy ion
collisions at the Relativistic Heavy Ion Collider \cite{MullerReview}, in
which the onset of a hydrodynamic regime is found to be earlier than weak
coupling estimates \cite{Son01}, and of describing the quantum quench that
indicates the unitary evolution of a quantum system with a sudden change of
coupling constants and can be realized experimentally in cold atom systems
\cite{Polkovnikov}.

In order to explore the dynamics and the scale dependence of thermalization
processes, the local observables, such as the expectation values of
energy-momentum tensor, can not provide sufficient information as in the
situation of viscous hydrodynamics. One important non-local observable is
the entanglement entropy (EE) between some spatial region and its complement
\cite{Nielsen00}. Usually EE is taken as a valuable probe to assess the
amount of entanglement and acts as an order parameter to witness various
quantum phases \cite{Wen0510}. Moreover, EE does not depend on the details
of theories and exists even in non-equilibrium quantum systems in which
there is no well-defined thermal entropy and temperature.

A precise holographic description of EE has been proposed via AdS/CFT
correspondence \cite{HEE}. It is calculated as the area of the minimum
surface in the bulk with its UV boundary coincident with the entangling
surface in the dual field theory. The prescription of holographic
entanglement entropy (HEE) has passed many nontrivial tests \cite{HEErev},
but there is no formal derivation until very recently \cite{Maldacena1304}.
Respecting the clear geometric image of HEE and the important role of
quantum entanglement in many-body quantum systems, the comparison between
HEE and EE may be very useful to provide new insights into the quantum
structure of spacetime \cite%
{Raamsdonk09,Swingle09,Vilaplana11,Bala1108,Takayanagi12,Takayanagi1302,Maldacena13,Myers1102,Hubeny1204,Hubeny1302}%
, particularly in the framework of the entanglement renormalization \cite%
{MERA}. Although HEE is previously defined only for static systems, a
covariant generalization is applicable to dynamical cases \cite{Hubeny0705},
where one should calculate HEE as the area of extremal surface and select
the minimum one if there are several extremal surfaces.

There is another interesting non-local observable, namely the mutual
information (MI) \cite{Nielsen00}, which measures the total (both classical
and quantum) correlations between two spatial subregions and acts as an
upper bound of the connected correlation functions in those regions \cite%
{Wolf07}. MI is related to EE closely. Considering two subsystems $A$ and $B$%
, one can define MI as $I(A,B)=S_{A}+S_{B}-S_{A\cup B}$, where $S_{A}$
denotes the EE on the subregion $A$. MI shares many features of EE in
nontrivial ways. For instance, EE has a ubiquitous area law, that is
divergent due to the presence of high energy singularities in unregulated
quantum field theories. The divergent contributions cancel in MI between
separate regions, leaving it as a scheme-independent quantity. But when $A$
and $B$ approach each other, the same short-distance divergence of EE
appears again \cite{Swingle1010}. It has been found in CFTs that MI has
power to extract more refined information than EE \cite{Furukawa08}. MI has
also been studied in strongly coupling field theories with gravity duals
both in static \cite{Headrick06,Hubeny0711,Swingle1010,Tonni1011,Vilaplana11}
and dynamic background \cite{BalaMI,Tonni11,Callan12}. In particular,\ for $%
l\gg \beta $ where $l$ is the size of $A$ and $\beta $ is the inverse
temperature, it was found \cite{Kundu12122} that the static HEE (both for $d$%
-dim relativistic CFTs and non-relativistic scale-invariant theories) can be
schematically decomposed as $S_{A}=S_{div}+S_{thermal}+S_{finite}+S_{corr}$,
where $S_{div}$ is the divergent boundary law, $S_{thermal}$ is thermal
entropy, $S_{finite}$ follows an area law and $S_{corr}$ is the correction
suppressed by exponentials of $l$. Accordingly, for $l\gg \beta $ and $x\ll
\beta $ (where $x$ is the size of the separation between two same subregions
$A$ and $B$), the holographic mutual information (HMI) can be decomposed as $%
I=I_{div}+S_{finite}+I_{corr}$, where $I_{div}=S_{div}$ appears in the limit
of $x\rightarrow 0$ and $I_{corr}$ are correction terms suppressed by
exponentials of $l$ and powers of $x$ \cite{Kundu12124}. Obviously, the
decomposition reveals that HMI can capture some important information of HEE{%
\footnote{%
In Ref. \cite{Kundu12124}, it was argued that HMI is a better guide than HEE
to capturing quantum entanglement.}}.

In this paper, we will investigate the entanglement entropy in the
thermalization process of the strongly coupled field theory following a
global quench. The holographic thermalization related to the global quench
have been discussed in \cite%
{Bhattacharyya0904,quench,Arrastia10,Aparicio11,Johnson11,Bala11,Keranen11,Galante1205,Hubeny1302,Liuwenbiao}%
. We will adapt the simple Vaidya model \cite%
{Arrastia10,Aparicio11,Johnson11,Bala11,Keranen11,Galante1205,Hubeny1302,Liuwenbiao}%
, which describes a homogeneous falling thin shell of null dust and is a
good quantitative approximation of the background generated by the
perturbation of a time-dependent scalar field \cite{Bhattacharyya0904} and
of the model of Ref. \cite{Lin08}. Among many interesting properties of
holographic thermalization that have been found in term of the Vaidya model,
it was observed that the evolution of HEE includes an intermediate stage
during which it is a simple linear function of time.{\ }The linear regime is
not obvious when $l$ is small, but it will be when $l$ increases. This
result matches well with the behavior seen in d=2 CFTs \cite{Cardy05,Cardy07}%
. Also it is consistent with the evolution of coarse-grained entropy in
nonlinear dynamical systems. There, it has been known that the linear growth
rate of coarse-grained entropy is generally described by the
Kolmogorov-Sinai entropy rate \cite{Latora99}. In classical 4-dim SU(2)
lattice gauge theory, the Kolmogorov-Sinai entropy rate is shown to be an
extensive quantity \cite{Muller00}. For strongly coupled field theory with
gravity dual, it has been found \cite{Bala11} that the growth rate of HEE
density in d=2 CFTs is also nearly volume-independent for small boundary
volumes. For large volumes, however, the growth rate of HEE approaches a
constant limit{\footnote{%
In Ref. \cite{Bala11}, the maximal growth rate is used to characterize the
linear growth, since the linear regime covers the time at which the growth
rate is maximal.}}.

One of the main motivations of this paper is to ask: whether the linear time
growth of HEE with a volume-independent rate is the dynamical correspondence
of the sub-leading area law of HEE captured by HMI in the static background?
More simply, can the dynamical HMI capture the constant growth rate? To
answer this question, we address the following work.

At first, we analytically prove that it is true for d=2 CFTs. Then we check
it using semi-analytic methods{\footnote{%
In this paper, we denote the numerical method as solving differential
equations numerically and the semi-analytic method as solving some
complicated algebra equations numerically.}}. Furthermore, by implementing a
time-consumed numerical computation, we can obtain the HEE with the large
enough volume to demonstrate the constant growth rate of dynamical HEE in
the d=3 relativistic CFT, the d=3 non-relativistic scale-invariant theory
and the dual theory of 5-dimensional (-dim) Gauss-Bonnet (GB) gravity. Also,
the linear growth with a constant slope is shown in the time evolution of
HMI. It deserves to note that we need to find a Vaidya metric in
asymptotically Lifshitz spacetime for studying the non-relativistic theory
and the formula of HEE of GB gravity has the nontrivial correction (not same
as the Wald entropy) to the one of Einstein gravity \cite%
{Fursaev06,GB1,Myers1101,Sinha1305}.

On the other hand, it is well known that the interior of black holes is
difficult to probe. So it is impressed that the holographic calculation of
dynamical non-local observables involves the information behind the apparent
horizon generated in the process of gravitational collapse \cite%
{Arrastia10,Aparicio11,Johnson11,Bala11,Asplund11,Basu12,Buchel13,BalaMI,Tonni11}%
. Recently, Hartman and Maldacena further isolated the origin of the linear
growth of HEE as arising from the growth of black hole interior measured
along a special critical spatial slice \cite{Maldacena13}. Their result was
obtained by studying the CFTs with the initial states of thermofield double
for thermal states and a particular pure state, which are dual to eternal
black holes and the eternal black holes with an end of the world brane that
cuts them in half, respectively.

Motivated by the insight that relating the horizon interior to the linear
growth of HEE, the second aim of this paper is to study whether the interior
of the apparent horizon along the extremal surface is also responsible for
the linear growth of HEE in\ Vaidya models{\footnote{%
It has been suggested that the apparent horizon may be more suitable than
the event horizon as a notion of entropy in the holographic theory out of
equilibrium \cite{Chesler0812,Hubeny0902,Hubeny0705,Takayanagi08}. With this
in mind, we prefer to demonstrate the relationship between apparent horizon
and the linear growth of HEE. But we will also study the event horizon in
the last of the paper.}}. We demonstrate that the answer is affirmative at
large $l$ in various holographic theories. Thus we can present a very
general result that in the process of holographic thermalization, the linear
growth with a constant rate of EE can be captured by MI in the boundary and
the horizon interior in the bulk.

The rest of the paper is arranged as follows. In Sec. II, we demonstrate
that the HEE has a linear growth regime when $l$ is large and the growth
rate approaches a constant limit when $l$ increases. In Sec. III, the
evolution of HMI is shown to contain the regime of the linear growth with
the constant rate. In Sec. IV, it is revealed that the extremal surfaces
probing the interior of apparent horizons account for the linear growth of
HEE at large $l$. In each section, we study the d=2, d=3 relativistic CFTs,
the d=3 non-relativistic scale-invariant theory and the dual theory of 5-dim
GB gravity, respectively. For d=2, we will use analytic and semi-analytic
methods. For other cases, only the numerical method is applicable. The
conclusion and discussion are given in Sec. V. We also add three appendix.
One is to look for a Vaidya metric in asymptotically Lifshitz spacetimes.
The second is to extend the decomposition of static HEE and HMI in Refs.
\cite{Kundu12122,Kundu12124} to the case of GB gravity by numerical fitting.
At last appendix, we study the extremal surface in the interior of event
horizon.

\section{Linear growth of HEE}

At the beginning, let us set a general frame that can accommodate all our
interested holographic theories as special cases. We will study the
thermalization processes of $d$-dim strongly coupled field theories modeled
by a homogeneous falling thin shell of null dust in $\left( d+1\right) $-dim
spacetime. Consider such a spacetime in the Poincar\`{e} coordinates%
\begin{equation}
ds^{2}=-\frac{1}{z^{2n}}f_{1}(z,v)dv^{2}-\frac{2}{z^{2}}f_{2}\left( z\right)
dzdv+\frac{1}{z^{2}L_{c}^{2}}d\vec{x}{}^{2}.  \label{Vaidya}
\end{equation}%
Here $z$ is the inverse of radial coordinate $r$. The spatial boundary
coordinates are denoted as $\vec{x}=(x_{1},\ldots ,x_{d-1})$ and the
translational invariance along $\vec{x}$ directions characterizes the global
quench in the boundary theory. In addition, $v$ labels the ingoing null
coordinate and we take the shell falling along $v=0$. The dynamical exponent
$n$ reflects the scale invariance and the effective curvature radius of
space can be given by $L_{c}^{2}=1/\left. f_{1}(z,v)\right\vert
_{z\rightarrow 0}$. For three kinds of holographic theories that we are
interested in{\footnote{%
We will denote them briefly as $d$-dim CFTs, Lifshitz gravity and GB\
gravity, respectively.}}, the unspecified quantities in the metric are
different but they are all restricted to be a Vaidya spacetime with a
massless shell. In addition, we will be interested in the case where the
shell is infalling and intermediates the vacuum and black brane.

For $d$-dim CFTs, the desired AdS-Vaidya collapse geometry has been
specified in \cite{Arrastia10,Bala11} with%
\begin{equation}
f_{1}(z,v)=1-m(v)z^{d},\;f_{2}\left( z\right) =1,\;n=1,\;L_{c}=1.
\label{CFTf}
\end{equation}%
Note that we set AdS radius as 1. The mass function of the shell is%
\begin{equation*}
m(v)=\frac{M}{2}\left[ 1+\tanh \frac{v}{v_{0}}\right] ,
\end{equation*}%
where $M$ denotes the mass for $v>v_{0}$ and $v_{0}$ represents a finite
shell thickness. We will be interested in the zero thickness limit, which
means to set the energy deposition on the boundary as instantaneous. Since
the general Vaidya metric for GB\ gravity has been found in \cite%
{Kobayashi05}, the asymptotically AdS geometry with null collapse is easily
obtained by requiring%
\begin{equation}
f_{1}(z,v)=\frac{1}{2\alpha }\left\{ 1-\sqrt{1-4\alpha \left[ 1-m(v)z^{d}%
\right] }\right\} ,\;f_{2}\left( z\right) =1,\;n=1,\;L_{c}=\sqrt{\frac{1+%
\sqrt{1-4\alpha }}{2}},  \label{GBf}
\end{equation}%
where $\alpha $ is the GB\ coupling constant. We also have interested in
studying the holographic thermalization of a 3-dim non-relativistic field
theory dual to a simple Lifshitz gravity \cite{Lif}, for which the static
HMI has been studied in \cite{Kundu12124}. So we construct an asymptotical
Lifshitz geometry with null collapse in Appendix A, which gives{\footnote{%
See other asymptotically-Lifshitz Vaidya spacetimes in \cite{Keranen11}.}}%
\begin{equation}
f_{1}(z,v)=1-m(v)z^{2},\;f_{2}\left( z\right) =z^{1-n},\;n=2,\;L_{c}=1.
\label{Lifshitzf}
\end{equation}

We will use HEE to probe the thermalization process. According to Ryu and
Takayanagi's proposition, the EE of a spatial region $A$ in a $d$-dim
strongly coupled field theory has a dual gravitational description, which
can be given by%
\begin{equation}
S=\frac{1}{4G_{N}^{d+1}}\int_{\Sigma }dx^{d-1}\sqrt{h},  \label{HEE}
\end{equation}%
where $G_{N}^{d+1}$ is the $\left( d+1\right) $-dim gravitational constant
and $h$ corresponds to the determinant of the induced metric of the minimal
surface $\Sigma $, which extends into the bulk and shares the boundary with $%
\partial A$. In dynamical cases, one should calculate HEE as the area of
extremal surface and select the minimal one if there are several extremal
surfaces. The prescription of Eq. (\ref{HEE}) has been used in
non-relativistic theories \cite{Solodukhin,Keranen11}. But for GB gravity,
it has been presented that Eq. (\ref{HEE}) should be modified as \cite%
{Fursaev06,GB1,Myers1101}%
\begin{equation}
S=\frac{1}{4G_{N}^{d+1}}\left\{ \int_{\Sigma }dx^{d-1}\sqrt{h}\left[ 1+\frac{%
2\alpha }{(d-2)(d-3)}\mathcal{R}_{\Sigma }\right] +\frac{4\alpha }{(d-2)(d-3)%
}\int_{\partial \Sigma }dx^{d-2}\sqrt{\sigma }\mathcal{K}\right\}
\label{HEEGB}
\end{equation}%
where $\mathcal{R}_{\Sigma }$ is the induced scalar curvature of surface $%
\Sigma $, $\sigma $ is the determinant of the induced metric of the boundary
$\partial \Sigma $, and $\mathcal{K}$ is the trace of the extrinsic
curvature of $\partial \Sigma $. Eq. (\ref{HEEGB}) should be extremized and
the minimal one should be selected as the definition of the HEE. Note that
the last term is the Gibbs-Hawking term that ensures a good variational
principle in extremizing the functional.

To fix the extremal surface, we need to specify the boundary region. In this
paper, we are interested in a rectangular boundary region with one dimension
of length $l$ and the other $d-2$ dimensions of volume $R^{d-2}$. We assume
that the rectangular strip is translationally invariant except along the $%
x_{1}$ direction. We also assume that $l$ is along $x_{1}$ direction and
denote $y=x_{1}$ for convenience.

In general, the extremal surfaces can be derived by extremizing Eq. (\ref%
{HEE}) or Eq. (\ref{HEEGB}). Substituting the Vaidya metric (\ref{Vaidya})
into them, HEE can be described as%
\begin{equation}
S=\frac{R^{d-2}}{4G_{N}^{d+1}}\int dy\frac{1}{z^{d-2}}\sqrt{%
z^{-2}-z^{-2n}f_{1}v^{\prime 2}-2z^{-2}f_{2}z^{\prime }v^{\prime }}
\label{HEE1}
\end{equation}%
for CFTs and Lifshitz gravity where $^{\prime }\equiv d/dy$, or%
\begin{equation}
S=\frac{R^{d-2}}{4G_{N}^{d+1}}\int \frac{dy}{d-2}\frac{z^{2}}{\left(
zL_{c}\right) ^{d+1}}L_{c}^{3}\Phi \Big\{(d-2)+2\alpha \Big[%
(d-4)z^{2}L_{c}^{2}+(d-2)\frac{z^{\prime 2}}{\Phi ^{2}}\Big]\Big\}
\label{HEE2}
\end{equation}%
for GB gravity with $\Phi =\sqrt{\frac{1}{L_{c}^{2}}-f_{1}v^{\prime
2}-2z^{\prime }v^{\prime }}$. Extremizing Eq. (\ref{HEE1}), one can derive
the two equations of motions%
\begin{eqnarray}
z^{\prime \prime }=\frac{z^{-4n}}{2f_{2}^{2}}\{z^{3}v^{\prime
2}f_{1}[2(d+n-2)f_{1}-z\partial _{z}f_{1}]-z^{2+2n}v^{\prime
}f_{2}(v^{\prime }\partial _{v}f_{1}+2z^{\prime }\partial _{z}f_{1})~ &&
\notag \\
+2~z^{1+2n}f_{1}[1-d+2(d+n-2)v^{\prime }z^{\prime }f_{2}]-2z^{4n}z^{\prime
2}f_{2}\partial _{z}f_{2}\}, &&  \label{eom1}
\end{eqnarray}%
\begin{equation}
v^{\prime \prime }=\frac{1}{2zf_{2}}\{2(d-1)(1-2v^{\prime }z^{\prime
}f_{2})+z^{2-2n}v^{\prime 2}[z\partial _{z}f_{1}-2(d+n-2)f_{1}]\}.
\label{eom2}
\end{equation}%
We will not write clearly the cumbersome equations of motions from Eq. (\ref%
{HEE2}).

To solve the equations of motion, one needs to fix the boundary conditions.
We set the two sides of the rectangular strip as $y=\pm \frac{l}{2}$ and set
boundary time $t_{0}=0$ when the shell just leaves the boundary. In summary,
the boundary conditions are%
\begin{equation}
z(\pm \frac{l}{2})=z_{0},\;v(\pm \frac{l}{2})=t_{0},  \label{boun1}
\end{equation}%
where $z_{0}$ is the cut-off close to the boundary. In addition, respecting
the symmetry of extremal surfaces in our setting, we have%
\begin{equation}
z^{\prime }(0)=0,\;v^{\prime }(0)=0.  \label{boun2}
\end{equation}%
Using these boundary conditions, one can try to solve the equations of
motions and obtain the HEE in terms of Eq. (\ref{HEE1}) and Eq. (\ref{HEE2}%
). However, the equations of motions are difficult to be solved analytically
in general.

At late time, the HEE $S$ will approach the equilibrium value $S_{thermal}$.
It can be obtained by Eq. (\ref{HEE1}) and Eq. (\ref{HEE2}) in the
background of pure black branes where the mass function $m(v)$ in $f_{1}$
should be replaced with the mass parameter $M$. Thus, one can obtain the
conservation equations%
\begin{equation}
1-z^{2-2n}f_{1}(z)v^{\prime 2}-2f_{2}(z)z^{\prime }v^{\prime }=\left( \frac{%
z_{\ast }}{z}\right) ^{2(d-1)}  \label{Con1}
\end{equation}%
for CFTs and Lifshitz gravity, and%
\begin{equation*}
\frac{1}{L_{c}}\Big(\frac{z_{\ast }}{z}\Big)^{d-1}\Big\{1-\frac{%
2(d-2)z^{\prime 2}L_{c}^{4}\alpha }{[d-2+2(d-4)z_{\ast }^{2}\alpha ]\Phi
_{1}^{2}}\Big\}=\Phi _{1},
\end{equation*}%
for GB gravity, where $z_{\ast }=z(0)$ and $\Phi _{1}=\sqrt{\frac{1}{%
L_{c}^{2}}-f_{1}(z)v^{\prime 2}-2z^{\prime }v^{\prime }}$. In terms of the
two conservation equations and the coordinate transformation $%
dv=dt-z^{n-1}dz/f_{1}$, the static HEE can be obtained. For CFTs and
Lifshitz gravity, it is given by
\begin{equation}
S_{thermal}=\frac{R^{d-2}}{2G_{N}^{d+1}}\int_{z_{0}}^{z_{\ast }}\frac{dz}{%
z^{d+2n-3}}\sqrt{\frac{2z^{3n-3}f_{2}(z)-1}{f_{1}(z)\left[ 1-(z/z_{\ast
})^{2(d-1)}\right] }}  \label{HEE3}
\end{equation}%
For GB\ gravity, the result is cumbersome and is not presented here clearly.

\subsection{d=2 CFTs}

In Ref. \cite{Bala11}, Balasubramanian et al. present an analytical method
to compute the length of geodesic in the AdS$_{3}$-Vaidya spacetime dual to
d=2 CFTs. The basic idea of this method is to separate the geodesic into the
part inside the shell and the part outside. The inside is described by the
pure AdS metric and the outside by the static BTZ black brane geometry.
Minimizing the total length of two parts of the geodesic, one can fix the
geodesic that is anchored at two sides of the rectangular strip on the
boundary and chases the shell falling along $v=0$. Let us review this method
briefly.

Outside the shell, the spacetime metric is%
\begin{equation*}
ds^{2}=-(r^{2}-r_{H}^{2})dt^{2}+\frac{dt^{2}}{r^{2}-r_{H}^{2}}+r^{2}dx^{2}%
\text{ with }t=v-\frac{1}{2r_{H}}\log \left\vert \frac{r-r_{H}}{r+r_{H}}%
\right\vert ,
\end{equation*}%
where $r_{H}$ is the location of horizon. The spatial geodesic is determined
by%
\begin{eqnarray}
\lambda _{\pm }^{out} &=&\frac{1}{2}\ln {\Big[-1+E^{2}-J^{2}+\frac{2r^{2}}{%
r_{H}^{2}}\pm \frac{2}{r_{H}^{2}}\sqrt{D(r)}\Big]},  \label{lamout} \\
t_{\pm }^{out} &=&t_{0}+\frac{1}{2r_{H}}\ln {\left\vert \frac{%
r^{2}-(E+1)r_{H}^{2}\pm \sqrt{D(r)}}{r^{2}+(E-1)r_{H}^{2}\pm \sqrt{D(r)}}%
\right\vert },  \notag \\
x_{\pm }^{out} &=&\frac{1}{2r_{H}}\ln {\Big[\frac{r^{2}-Jr_{H}^{2}\pm \sqrt{%
D(r)}}{r^{2}+Jr_{H}^{2}\pm \sqrt{D(r)}}\Big]},  \label{xrout} \\
v_{\pm }^{out} &=&t_{0}+\frac{1}{2r_{H}}\ln {\Big[\frac{r-r_{H}}{r+r_{H}}%
\frac{r^{2}-(E+1)r_{H}^{2}\pm \sqrt{D(r)}}{r^{2}+(E-1)r_{H}^{2}\pm \sqrt{D(r)%
}}\Big]},  \label{vout} \\
D(r) &=&r^{4}+(-1+E^{2}-J^{2})r_{H}^{2}r^{2}+J^{2}r_{H}^{4},  \notag
\end{eqnarray}%
where $E$ and $J$ are conserved charges concerning energy and angular
momentum, respectively. The subscript \textquotedblleft $+$%
\textquotedblright\ denotes branch 1 and \textquotedblleft $-$%
\textquotedblright\ means branch 2. Both of them are necessary to give the
complete geodesic in general. The superscript \textquotedblleft $out$%
\textquotedblright\ denotes the part of the geodesic outside the shell. The
part inside the shell, that is denoted by the superscript \textquotedblleft $%
in$\textquotedblright , can be described similarly by
\begin{eqnarray}
\lambda _{\pm }^{in} &=&\pm \cosh ^{-1}(\frac{r}{r_{\ast }}),  \label{lamin}
\\
t^{in} &=&\frac{1}{r_{sw}}=const.,  \notag \\
x_{\pm }^{in} &=&\pm \frac{1}{r_{\ast }}\sqrt{1-\left( \frac{r_{\ast }}{r}%
\right) ^{2}},  \label{xrin} \\
v^{in} &=&\frac{1}{r_{sw}}-\frac{1}{r},  \label{vin}
\end{eqnarray}%
where $r_{\ast }$ denotes the radial endpoint of the geodesic inside the
shell and $r_{sw}$ is the radial location at which the geodesic intersects
the shell. Here \textquotedblleft $\pm $\textquotedblright\ denote that the
part of geodesic in pure AdS region is symmetric. Extremizing the total
geodesic length, one can obtain the refraction conditions, which give out
the conserved charges of the outside geodesic. For branch 1 with $r_{sw}\geq
r_{H}/\sqrt{2}$ and branch 2 with $r_{sw}\leq r_{H}/\sqrt{2}$, the conserved
charges are%
\begin{equation}
E=-\frac{r_{H}\sqrt{r_{sw}^{2}-r_{\ast }^{2}}}{2r_{sw}^{2}},\ J=+\frac{%
r_{\ast }}{r_{H}},  \label{EJ}
\end{equation}%
where we have selected the proper sign combination to ensure a finite $v$
when the geodesic crosses the future horizon. With the mind that the shell
is falling along $v=0$ and the spatial separation on the boundary is fixed
as $l$, one can obtain the parameters $r_{sw}$ and $r_{\ast }$ from%
\begin{eqnarray}
2\rho &=&\coth \left( {r_{H}t_{0}}\right) +\sqrt{\coth ^{2}\left( {r_{H}t_{0}%
}\right) -\frac{2c}{c+1}}  \label{ro} \\
l &=&\frac{1}{r_{H}}\left\{ \frac{2c}{s\rho }+\ln \left[ \frac{2\left(
1+c\right) \rho ^{2}+2s\rho -c}{2\left( 1+c\right) \rho ^{2}-2s\rho -c}%
\right] \right\} ,  \label{l}
\end{eqnarray}%
where%
\begin{equation}
\rho =r_{sw}/r_{H},\;\rho s=r_{\ast }/r_{H},\;s=\sqrt{1-c^{2}}.  \label{rocs}
\end{equation}%
Finally, the sum of the length of inside and outside geodesics can be
written as%
\begin{eqnarray}
L(l,t_{0}) &=&2\left[ \lambda _{+}^{in}(r_{sw})-\lambda _{+}^{in}(r_{\ast })%
\right] +2\left[ \lambda _{+}^{out}(r_{0})-\lambda _{-}^{out}(r_{sw})\right]
\label{L1} \\
&=&2\ln \left[ \frac{2r_{0}\sinh \left( r_{H}t_{0}\right) }{r_{H}s\left(
l,t_{0}\right) }\right] ,  \label{L2}
\end{eqnarray}%
where $r_{0}=1/z_{0}$ denotes a UV cutoff and $s\left( l,t_{0}\right) $ is
an implicit function\ determined by Eqs. (\ref{ro}), (\ref{l}) and (\ref%
{rocs}). In Eq. (\ref{L1}), it seems that we have assumed the branch 2
intermediates the branch 1 and the inside geodesic. However, the result is
same for another case in which the branch 1 connects the inside geodesic
directly. This is because $\lambda _{+}^{out}(r_{sw})$ with $r_{sw}\geq
r_{H}/\sqrt{2}$ has the same form as $\lambda _{-}^{out}(r_{sw})$ with $%
r_{sw}\leq r_{H}/\sqrt{2}$.

Although the geodesic has been described by analytical formula, the implicit
function $s\left( l,t_{0}\right) $ in Eq. (\ref{L2}) can be solved only by
numerical methods in general. This is why we call this method as the
semi-analytical method. Fortunately, the analytical expansion of Eq. (\ref%
{L2}) has been found in the large boundary region \cite{Aparicio11} or in
the period of the early time growth and late time saturation \cite%
{Hubeny1302}. They are even applicable to calculate the non-equal time
two-point functions and allow for the different geometry inside the shell.
Here we will give a simplified version of the analytical method which is
enough for giving an analytical solution in the region that we have
interested in, namely, the large $lr_{H}$ and intermediate $t_{0}r_{H}$
(Hereafter, we will set $r_{H}=1$ for convenience when we discuss the region
of parameters sometimes and plot all the figures. But we keep it clear in
all the formula.).

\begin{figure}[tbp]
\centering
\includegraphics[width=1\textwidth]{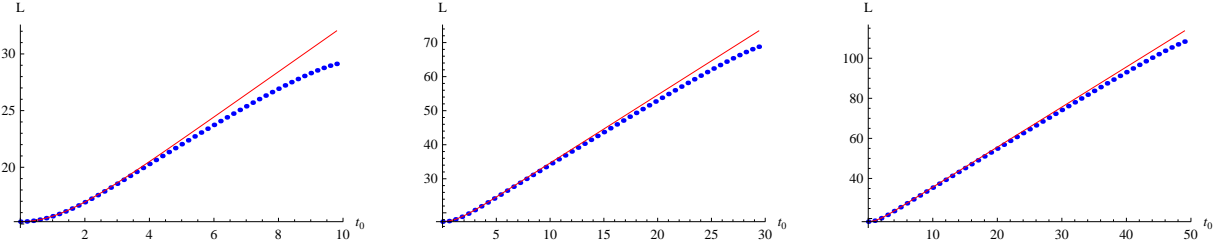}
\caption{Compare the geodesic length $L(l,t_{0})$ with analytical and
semi-analytical formula, which are expressed using the red lines and blue
points, respectively. The left, middle and right panels use the boundary
separations $l=20,60,100\,$, respectively. The UV cutoff has been set as $%
r_{0}=100$.}
\label{L}
\end{figure}

Note a simple but important observation from Eqs. (\ref{ro}), (\ref{l}) and (%
\ref{rocs}), that is, the implicit function $s\left( l,t_{0}\right) \in %
\left[ 0,1\right] $ and it decreases when $l\rightarrow \infty $ or $%
t_{0}/l\rightarrow 0$. Thus, we can expand Eq. (\ref{l}) as%
\begin{equation*}
l=\frac{4\tanh \left( \frac{r_{H}t_{0}}{2}\right) }{r_{H}s}+\mathcal{O}%
(s)^{1}.
\end{equation*}%
Immediately, one can have{\footnote{%
After finishing this paper, we were kindly informed by E. Lopez that Eq. (%
\ref{S1}) has been obtained in \cite{Aparicio11}.}}%
\begin{equation}
L(l,t_{0})=2\ln \left[ lr_{0}\cosh ^{2}\left( \frac{r_{H}t_{0}}{2}\right) %
\right] .  \label{S1}
\end{equation}%
To see the effectiveness of Eq. (\ref{S1}), we compare it with the
semi-analytical result of Eq. (\ref{L2}) in Fig. \ref{L}.
One can find that they match well in a larger region of $t_{0}$ when $%
l\rightarrow \infty $. Furthermore, the derivative of Eq. (\ref{S1}) is%
\begin{equation}
\frac{dL}{dt_{0}}=2r_{H}\tanh \left( \frac{r_{H}t_{0}}{2}\right) ,
\label{dLdt01}
\end{equation}%
which approaches a constant limit fast when $t_{0}$ increases, see Fig. \ref%
{dLdt0}.
\begin{figure}[tbp]
\centering
\includegraphics[width=0.5\textwidth]{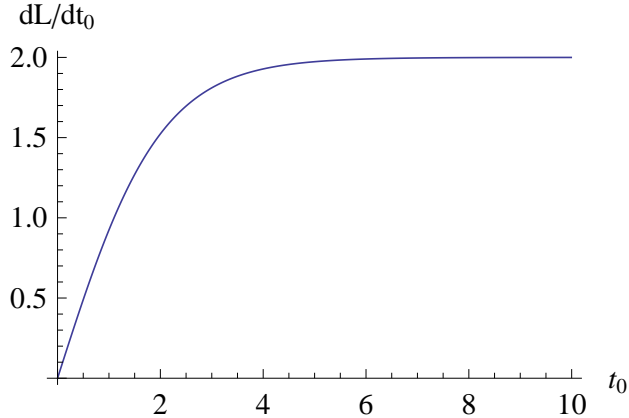}
\caption{The derivative of geodesic length $L$ with respect to $t_{0}$ as a
function of $t_{0}$, using the analytical method.}
\label{dLdt0}
\end{figure}

\subsection{d=3 CFTs}

For other theories with $d>2$, we will use numerical methods to solve two
equations of motion Eq. (\ref{eom1}) and Eq. (\ref{eom2}) with boundary
conditions Eq. (\ref{boun1}) and Eq. (\ref{boun2}). Here we fix two small
parameters as $v_{0}=0.01$ and $z_{0}=0.01$. In order to obtain the
numerical solutions, it would be found that the precision and time of the
computation increase fast when $t_{0}$ and $l$ increase. Fortunately, since
we expect that the linear growth of HEE appears in the intermediate $t_{0}$,
it is not necessary to solve the equations of motion in the region of very
large $t_{0}$.

For d=3 CFTs, substituting Eq. (\ref{CFTf}) into equations of motion and
implementing a time-consumed computation with high precision, we can obtain $%
z(y)$ and $v(y)$ with fixed $t_{0}$ and $l$. Consequently, we can integrate
Eq. (\ref{HEE1}), which shows clearly that in the region of the intermediate
$t_{0}$ and large $l$, HEE grows linearly{\footnote{%
For the holographic theories with $d>2$, there would be two or three
extremal surfaces when $t_{0}$ increases. We have selected the one with the
minimal HEE.}} and the growth rate approaches a constant when $l$ increases,
see Fig. \ref{CFT3HEE}. Note that we are comparing the HEE at any given time
with the late time result $S_{thermal}$, which can be obtained from Eq. (\ref%
{HEE3}).{\
\begin{figure}[tbp]
\includegraphics[width=1\textwidth]{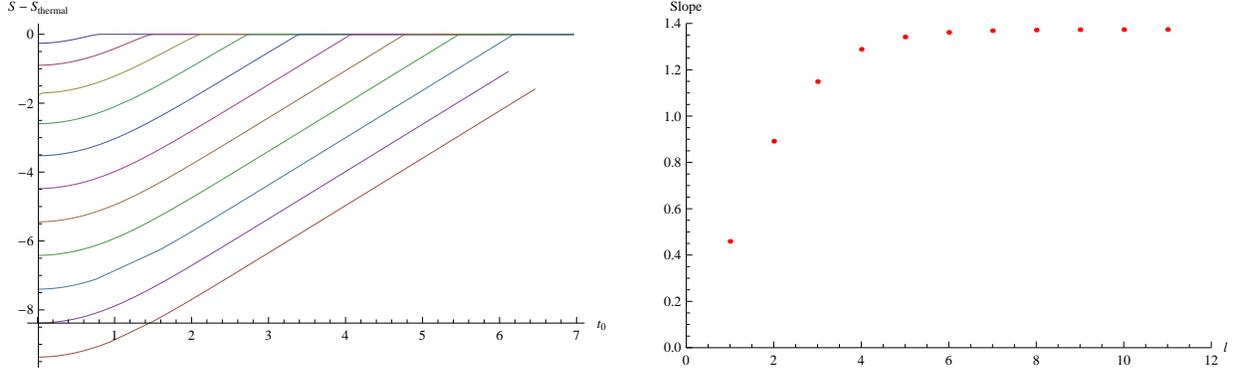}\newline
\caption{Left: $S-S_{thermal}$ for d=3 CFTs as a function of $t_{0}$ with
different $l$ from 1 (top) to 11 (down). Right: The derivative of HEE with
respect to $t_{0}$ at an intermediate value $t_{0}=l/2$. The behaviour of
the derivative will be not qualitatively changed if one selects other $t_{0}$%
, provided that it is not too small or too close to the thermalization time
(that is what we mean by the \textquotedblleft
intermediate\textquotedblright\ time). We set $4G_{N}^{d+1}=R^{d-2}=1$ in
all the figures for convenience.}
\label{CFT3HEE}
\end{figure}
}

\subsection{Lifshitz gravity}

\begin{figure}[tbp]
\includegraphics[width=1\textwidth]{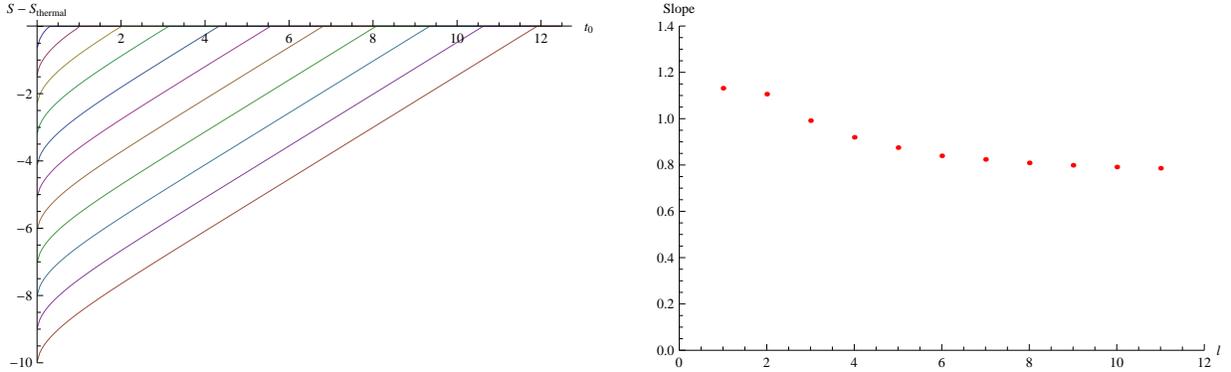}\newline
\caption{Left: $S-S_{thermal}$ for Lifshitz gravity as a function of $t_{0}$
with different $l$ from 1 (top) to 11 (down). Right: The derivative of HEE
with respect to $t_{0}$ at an intermediate value $t_{0}=l/4$.}
\label{LifHEE}
\end{figure}

Next we will consider the Lifshitz background, which can be regarded as the
holographic dual to the non-relativistic scale-invariant (non-conformal)
field theory. Solving the equations of motion with Eq. (\ref{Lifshitzf}) and
integrating Eq. (\ref{HEE1}), one can see the time evolution of HEE with
different $l$ in Fig. \ref{LifHEE}.

\subsection{GB gravity}

In terms of Eq. (\ref{GBf}) and Eq. (\ref{HEE2}), the similar behavior can
be found in the HEE of field theories dual to 5-dim GB gravity, see Fig. \ref%
{GBHEE}.
\begin{figure}[tbp]
\includegraphics[width=1\textwidth]{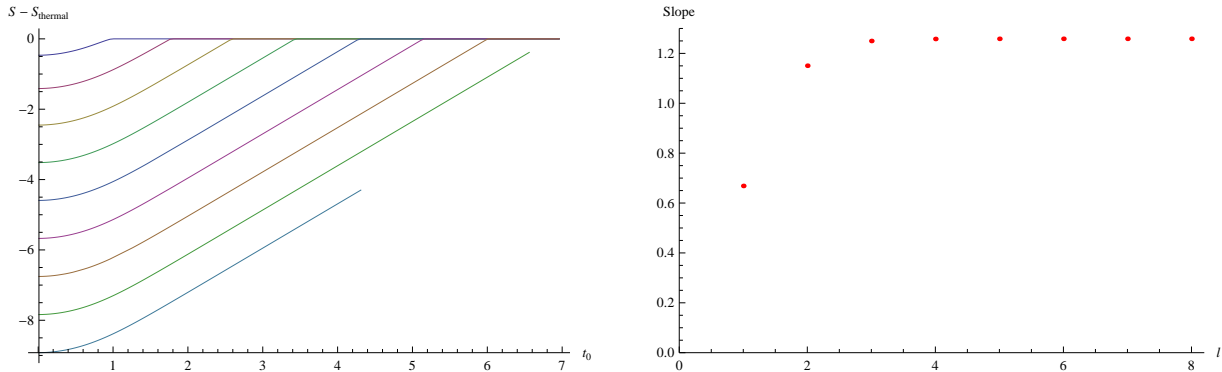}\newline
\caption{Left: $S-S_{thermal}$ for GB gravity as a function of $t_{0}$ with
different $l$ from 1 (top) to 9 (down). Right: The derivative of HEE with
respect to $t_{0}$ at an intermediate value $t_{0}=2l/3$. We set $\protect%
\alpha =0.05$ hereafter.}
\label{GBHEE}
\end{figure}

\section{Linear growth of HMI}

By decomposing HEE and HMI \cite{Kundu12122,Kundu12124}, Fischler et al.
found that, when $l\gg \beta $, HMI contains the sub-leading area law of HEE
in the static background. It was also shown that the decomposition of HEE
and HMI is general for both $d$-dim relativistic CFTs and non-relativistic
scale-invariant theories. In Appendix B, we prove that there is a similar
decomposition in the holographic theories dual to GB gravity. Respecting the
nontrivial correction of GB effect to the prescription of HEE, we believe
that the decomposition is a very general result in strongly coupling field
theories.

In this section, we will investigate the dynamical behavior of HMI, focusing
on the region of large $l$. Our aim is to prove that the dynamical HMI at
large $l$ also can capture the behavior of the linear time growth with the $%
l $-independent rate of dynamical HEE that was shown in the above section.
Let us introduce the prescription of HMI. Consider two disjoint rectangular
subregion $A$ and $B$. We set that they are same with one dimension of
length $l$ and are separated with distance $x$. HMI is defined by HEE as%
\begin{equation*}
I(A,B)=S_{A}+S_{B}-S_{A\cup B}\text{.}
\end{equation*}%
For $x\neq 0$, there may be three choices of extremal surfaces which are
anchored on the boundary of $A\cup B$ \cite{BalaMI}. But since in all the
holographic theories we have shown that the HEE is monotonically increasing
with respect to $l$, it is enough to consider HMI as%
\begin{equation}
I(A,B)=2S(l,t_{0})-\min [2S(l,t_{0}),S(2l+x,t_{0})+S(x,t_{0})],  \label{I1}
\end{equation}%
where $S(l,t_{0})$ denotes the HEE on the region $A$ (or $B$).

\subsection{d=2 CFTs}

\subsubsection{Analytical method for HMI with small x}

Here we will study the HMI with large $l$ based on the analytical expression
of HEE Eq. (\ref{S1}). Obviously, Eq. (\ref{S1}) is not applicable to
compute the HMI with general $x$. But fortunately, we can compute it for
large $l$, intermediate $t_{0}$ and small $x$. This is because the troubled
term $S(x,t_{0})$ in Eq. (\ref{I1}) achieves the equilibrium value when $%
t_{0}>x/2$ \cite{Arrastia10} and can be replaced with the static one%
\begin{equation*}
S(x)=\frac{1}{2G_{N}^{3}}\ln \left[ \frac{2r_{0}\sinh \left( \frac{r_{H}l}{2}%
\right) }{r_{H}}\right] .
\end{equation*}%
Thus, for $x\ll \beta \ll l$, Eq. (\ref{I1}) reads as%
\begin{eqnarray*}
I &=&2S(l,t_{0})-\min [2S(l,t_{0}),S(2l+x,t_{0})+S(x)] \\
&=&\frac{1}{2G_{N}^{3}}\ln \left[ \frac{l}{2x}\cosh ^{2}\left( \frac{%
r_{H}t_{0}}{2}\right) \right] +\mathcal{O}(\frac{x}{l})+\mathcal{O}%
(xr_{H})^{2}.
\end{eqnarray*}%
Note that $S(2l+x,t_{0})+S(x)$ is smaller than $2S(l,t_{0})$ when $x$ is so
small. Comparing the dominated term with Eq. (\ref{S1}), one can find that
HMI contains the exact time-dependent part of HEE in the region of small $x$.

\subsubsection{Semi-analytical method for HMI with general x}

\begin{figure}[tbp]
\includegraphics[width=1\textwidth]{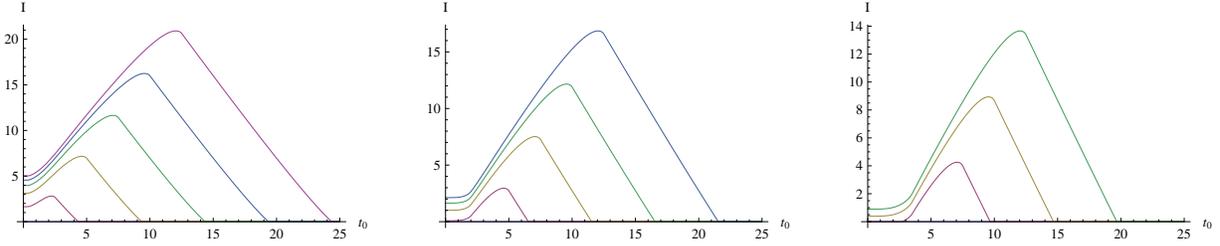}\newline
\caption{HMI for d=2 CFTs as a function of $t_{0}$, using the
semi-analytical method. From top to down, the boundary separations were
taken as $l=25,20,15,10,5,2$. But HMI vanishes for $l=2$ in the left panel
with $x=1$, for $l=2,5$ in the middle panel with $x=4$, and for $l=2,5,10$
in the right panel with $x=7$.}
\label{HMI1}
\end{figure}

\begin{figure}[tbp]
\includegraphics[width=1\textwidth]{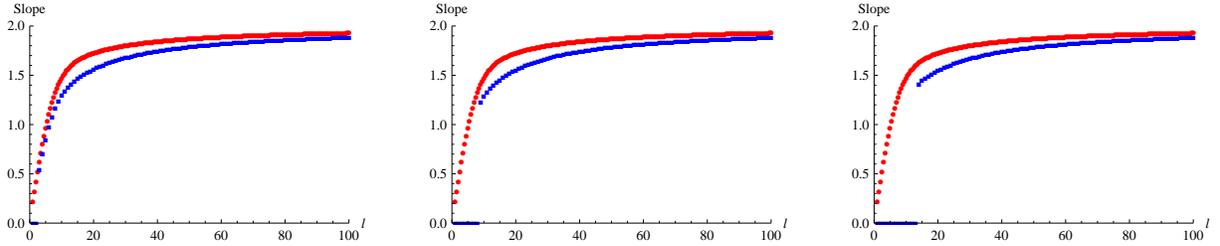}\newline
\caption{The growth rate of HEE (red points) for d=2 CFTs and HMI (blue
points) at $t_{0}=l/4$ as a function of $l$ and $x=1,4,7$ for left, middle
and right panels. The nonvanishing part of blue points at large $l$ is
nearly independent with $x$.}
\label{DIS}
\end{figure}

Using the semi-analytical method, we can study HMI in the complete region of
all the parameters, see Fig. \ref{HMI1}{\footnote{%
Note that the dynamical HMI in $d=2$ and $d=3$ CFTs have been studied in
Refs. \cite{BalaMI,Tonni11}, but they did not pay attention to the constant
growth rate that appears in the region of large $l$.}}.
In this figure, we will not care about the region with $t_{0}>l/2$ and $%
t_{0}>x/2$, where $S(l,t_{0})$ and $S(x,t_{0})$ have achieved the
equilibrium and HMI trivially reflects the dynamical behavior of $%
S(2l+x,t_{0})$. Instead, we focus on the region with $t_{0}<l/2$. One can
find that the slope of the linear growth of HMI approaches a constant as $l$
increases. To compare the slope of HMI and HEE, we plot in Fig. \ref{DIS}
the derivative of HMI\ and HEE with respect to $t_{0}$ at an intermediate
value of $t_{0}=l/4$ during the linear growth period.
From this figure, one can find that the linear growth rate of HMI with
different $x$ approaches the rate of HEE when $l$ increases. The effect of
increasing $x$ only enlarges the vanishing region of the growth rate of HMI
at small $l$ but the nonvanishing part at large $l$ is nearly independent
with $x$. Thus, we have shown that HMI can capture the behavior of the
linear growth of the HEE with the $l$-independent rate even for general $x$.

\subsection{Other holographic theories}

\begin{figure}[tbp]
\includegraphics[width=1\textwidth]{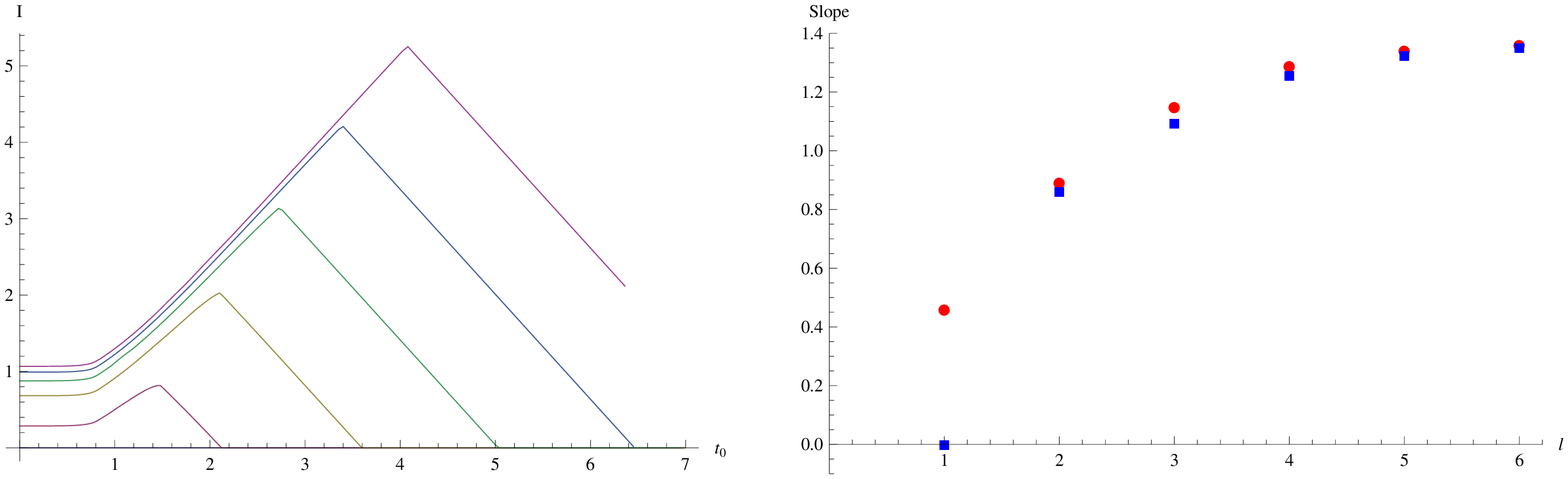}\newline
\caption{(Left) HMI for d=3 CFTs as a function of $t_{0}$. The boundary
separations were taken as $l=6$ (top) to 1 (down). HMI vanishes for $l=1$.
(Right) The growth rate of HEE (red points) and HMI (blue points) at $%
t_{0}=l/2$ as a function of $l$. We set $x=1$.}
\label{DISd3CFT}
\end{figure}

\begin{figure}[tbp]
\includegraphics[width=1\textwidth]{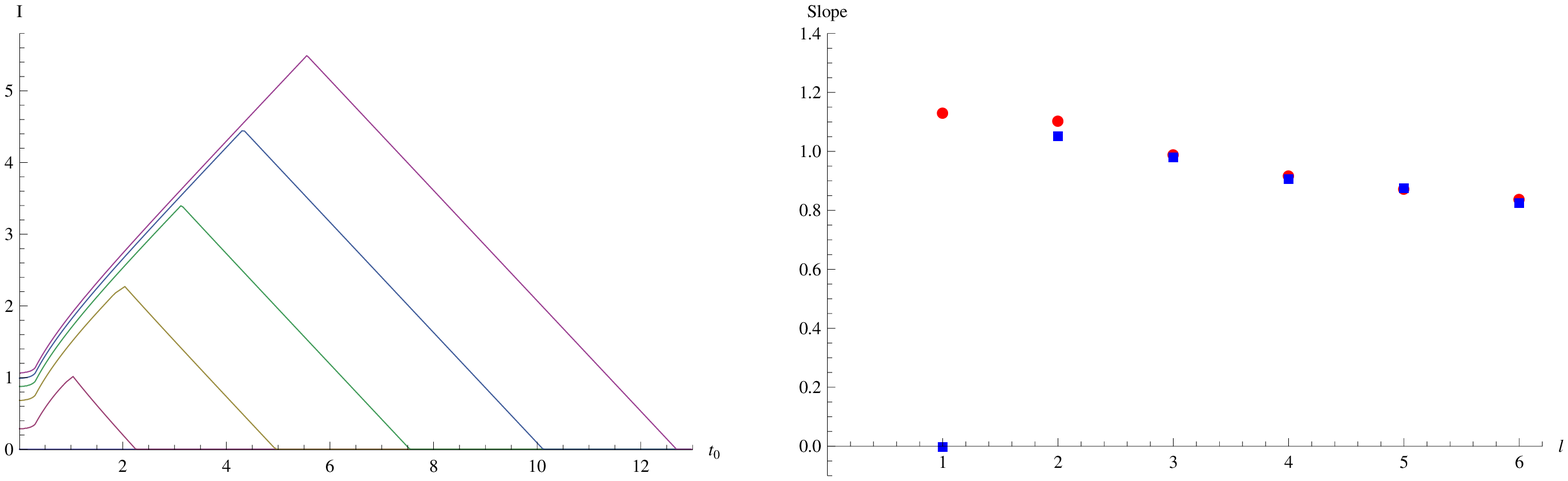}\newline
\caption{(Left) HMI for Lifshitz gravity as a function of $t_{0}$. The
boundary separations were taken as $l=6$ (top) to 1 (down). HMI vanishes for
$l=1$. (Right) The growth rate of HEE (red points) and HMI (blue points) at $%
t_{0}=l/4$ as a function of $l$. We set $x=1$.}
\label{DISLif}
\end{figure}

\begin{figure}[tbp]
\includegraphics[width=1\textwidth]{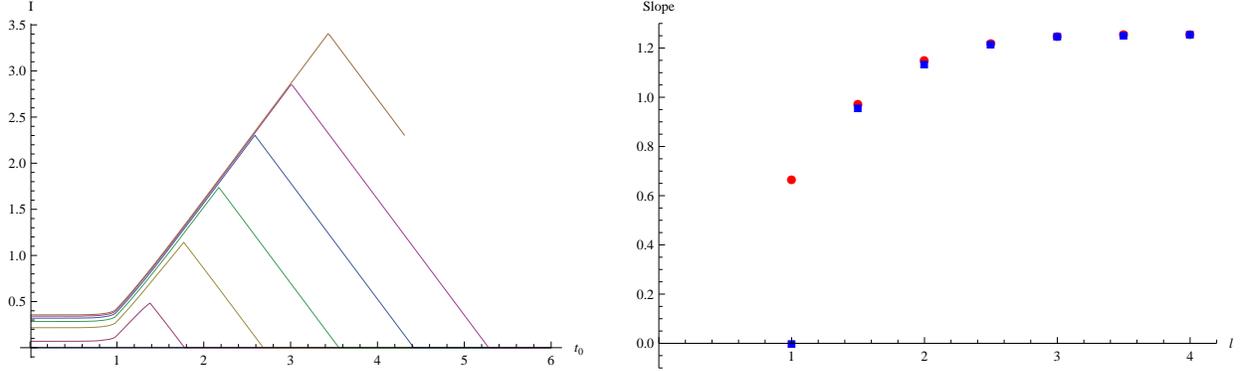}\newline
\caption{(Left) HMI for GB gravity as a function of $t_{0}$. The boundary
separations were taken as $l=4,3.5,3,2.5,2,2.5,1$ from top to down. HMI
vanishes for $l=1$. (Right) The growth rate of HEE (red points) and HMI
(blue points) at $t_{0}=2l/3$ as a function of $l$. We set $x=1$.}
\label{DISGB}
\end{figure}

Using the numerical method, we study the HMI in different theories. From
Fig. \ref{DISd3CFT} to Fig. \ref{DISGB},
one can find that the linear time growth and the constant rate of dynamical
HEE are presented in HMI for all the cases.

\section{Linear growth of HEE from the interior of apparent horizons}

Motivated by Hartman and Maldacena's work \cite{Maldacena13}, we will study
the relationship between the linear growth of HEE in Vaidya models and the
extension of the extremal surface in the interior of the apparent horizon
(Note that by \textquotedblleft interior\textquotedblright , it means the
region between the location of the apparent horizon and the singularity).
Let us introduce the apparent horizon. It is sometimes called as marginal
surfaces, defined as the boundary of trapped surfaces associated to a given
foliation \cite{Hawking73}. Thus, its location can be determined by one
vanishing null expansion. In terms of the general metric (\ref{Vaidya}), the
tangent vector of ingoing and outgoing radial null geodesics can be read as%
\begin{equation}
N^{in}=\frac{z^{2}}{f_{2}(z)}\partial _{z},\;N^{out}=\partial _{v}-\frac{1}{%
2z^{2n-2}}\frac{f_{1}(z,v)}{f_{2}\left( z\right) }\partial _{z},  \label{NN}
\end{equation}%
where we have used the normalization $N^{in}\cdot N^{out}=-1$. The expansion
along outgoing null geodesics is given by%
\begin{equation}
\theta =P^{\mu \nu }\nabla _{\mu }N_{v}^{out}  \label{th}
\end{equation}%
with the projective tensor%
\begin{equation*}
P_{\mu \nu }=g_{\mu \nu }+N_{\mu }^{in}N_{v}^{out}+N_{v}^{in}N_{\mu }^{out}.
\end{equation*}%
Using Eqs. (\ref{Vaidya}), (\ref{NN}) and (\ref{th}) we have%
\begin{equation*}
\theta =\frac{d-1}{2z^{2n-1}}\frac{f_{1}(z,v)}{f_{2}\left( z\right) }.
\end{equation*}%
Thus, for the holographic theories that we are interested in, the location
of apparent horizons $r_{A}\left( v\right) $ is determined by $f_{1}(z,v)=0$.

\subsection{d=2 CFTs}

\subsubsection{Analytical method for the geodesic inside the horizon}

\begin{figure}[tbp]
\includegraphics[width=1\textwidth]{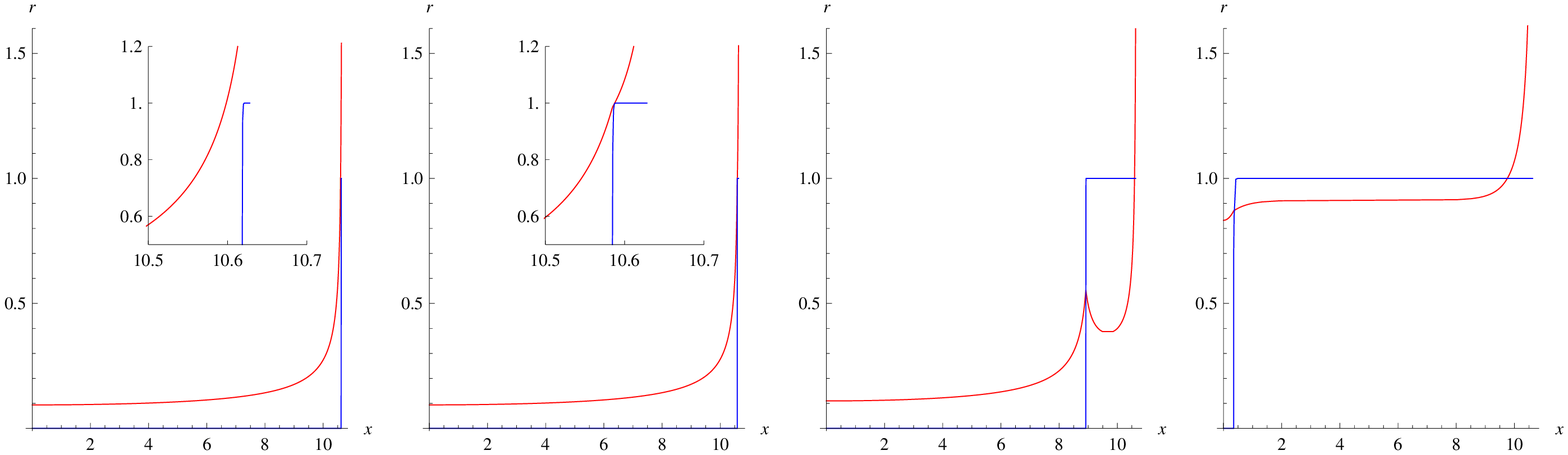}\newline
\caption{Four typical relation between geodesics (red) and apparent horizons
(blue) in the $(x,r)$ plane. $l$ is fixed as 21.3. From left to right, $%
t_{0}=0.8,1.12,4,10.6$.}
\label{xr}
\end{figure}

Now we will use the analytical description of the geodesic in Sec. II. A to
isolate the part of HEE contributed by the geodesic in the interior of
apparent horizons. For this aim, let us plot some typical geodesics and
apparent horizons using Eqs. (\ref{xrout}), (\ref{vout}), (\ref{xrin}) and (%
\ref{vin}), see Fig. \ref{xr}.
From these curves, one can extract three facts. First, it is possible that
the geodesic crosses the apparent horizon twice. One crosspoint $p$ is
located at $r_{p}=r_{H}$ and the other $q$ at $r_{q}=r_{sw}$. Note that the
location of crosspoint $q$ can be understood since $r_{A}\left( v\right) $
is a step function vanishing at $v<0$ and the nonvanishing radial location
of crosspoints in Fig. \ref{xr} except $r_{p}=r_{H}$ should be located at
the position with $v(r_{sw})=0$. Second, when a geodesic crosses the
apparent horizon, the branch 1 may crosses the horizon twice (see the
rightmost panel in Fig. \ref{xr}) or the branch 1 crosses at $r_{p}=r_{H}$
and the branch 2 crosses at $r_{q}=r_{sw}$ (see the third panel from left in
Fig. \ref{xr}). But as we have mentioned below Eq. (\ref{L2}), it is not
necessary to consider both cases in the calculation since $\lambda
_{+}^{out}(r_{sw})$ with $r_{sw}\geq r_{H}/\sqrt{2}$ has the same form as $%
\lambda _{-}^{out}(r_{sw})$ with $r_{sw}\leq r_{H}/\sqrt{2}$. Third, the
geodesic does not cross the apparent horizon when $r_{sw}>r_{H}$. With these
facts in mind, we can write the length of the geodesic between two
crosspoints as%
\begin{equation}
L_{interior}(l,t_{0})=2\left[ \lambda _{+}^{out}(r_{H})-\lambda
_{-}^{out}(r_{sw})\right] \text{, with }r_{sw}(l,t_{0})<r_{H}  \notag
\end{equation}%
\begin{equation}
=\ln \left[ \frac{4r_{sw}^{4}\left( r_{H}^{2}-r_{\ast }^{2}\right)
+4r_{H}^{3}r_{sw}^{2}\sqrt{r_{sw}^{2}-r_{\ast }^{2}}+r_{H}^{4}\left(
r_{sw}^{2}-r_{\ast }^{2}\right) }{r_{H}^{4}\left( r_{sw}^{2}-r_{\ast
}^{2}\right) -4r_{H}^{2}r_{sw}^{3}\left( r_{sw}+\sqrt{r_{sw}^{2}-r_{\ast
}^{2}}\right) -4r_{sw}^{4}\left[ r_{\ast }^{2}-2r_{sw}\left( r_{sw}+\sqrt{%
r_{sw}^{2}-r_{\ast }^{2}}\right) \right] }\right] ,  \label{Linterior}
\end{equation}%
where we have used the outside solutions of geodesic Eq. (\ref{lamout}) and
neglected the clear expression of restriction $r_{sw}(l,t_{0})<r_{H}$
hereafter for convenience. In terms of Eq. (\ref{rocs}), we expand Eq. (\ref%
{Linterior}) with respect to $s$%
\begin{equation}
L_{interior}(l,t_{0})=2\ln \left[ \frac{e^{r_{H}t_{0}}-1}{2}\right] +%
\mathcal{O}(s)^{2}.  \label{Linterior1}
\end{equation}%
The derivative of its dominated term is%
\begin{equation}
\frac{dL_{interior}}{dt_{0}}=\frac{2r_{H}}{1-e^{-r_{H}t_{0}}}.
\label{Ldt0interior}
\end{equation}%
From Eq. (\ref{Linterior1}) and Eq. (\ref{Ldt0interior}), it is explicit
that the length $L_{interior}$ grows linearly and the slope approaches the
constant $2r_{H}$ fast when $t_{0}$ increases, which is exactly same as the
behavior of HEE seen in Eq. (\ref{S1}) and Eq. (\ref{dLdt01}). To be more
clear, we also compute the difference between Eq. (\ref{S1}) and Eq. (\ref%
{Linterior1})%
\begin{equation*}
L(l,t_{0})-L_{interior}(l,t_{0})=2\ln [\frac{%
lr_{0}e^{-r_{H}t_{0}}(1+e^{r_{H}t_{0}})^{2}}{2\left( e^{r_{H}t_{0}}-1\right)
}],
\end{equation*}%
which is close to the constant $2\log (lr_{0}/2)$ when $t_{0}$ increases.
Thus, we have proven analytically that the linear growth of HEE in the
region of large $l$ and intermediate $t_{0}$ completely comes from the
growth of geodesic length inside the apparent horizon.

\subsubsection{Semi-analytical method for the geodesic inside the horizon}

\begin{figure}[tbp]
\includegraphics[width=1\textwidth]{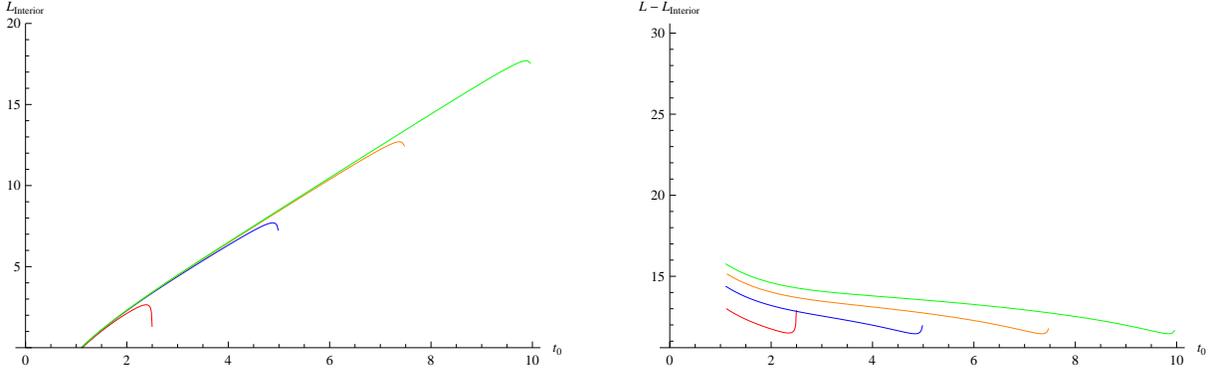}\newline
\caption{$L_{interior}$ (left) and $L-L_{interior}$ (right) for d=2 CFTs as
functions of $l$ and $t_{0}$. The boundary separations were taken to be $%
l=5,10,15,20$ (red, blue, orange, green).}
\label{LjL}
\end{figure}

Based on Eq. (\ref{Linterior}) and the numerical solution of the implicit
function $s(l,t_{0})$, we can plot $L_{interior}(l,t_{0})$ in the complete
region of parameters, see Fig. \ref{LjL}.
One can find that the length $L_{interior}$ grows linearly and its
difference with $L$ approaches a constant in the region of large $l$ and
intermediate $t_{0}$. The result is consistent with the analytical method,
as it should be.

\subsection{Other holographic theories}

\begin{figure}[tbp]
\includegraphics[width=1\textwidth]{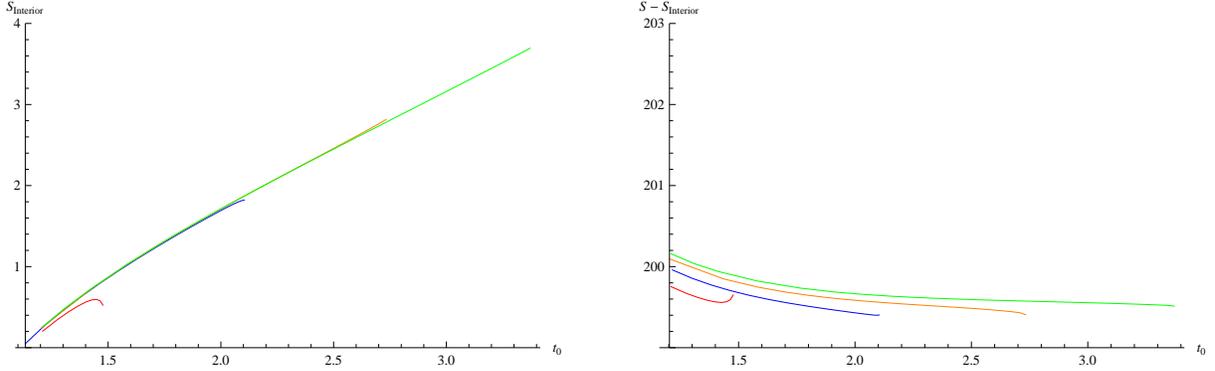}\newline
\caption{$S_{interior}$ (left) and $S-S_{interior}$ (right) for d=3 CFTs as
functions of $l$ and $t_{0}$. The boundary separations were taken to be $%
l=2,3,4,5$ (red, blue, orange, green).}
\label{LjLd3}
\end{figure}

\begin{figure}[tbp]
\includegraphics[width=1\textwidth]{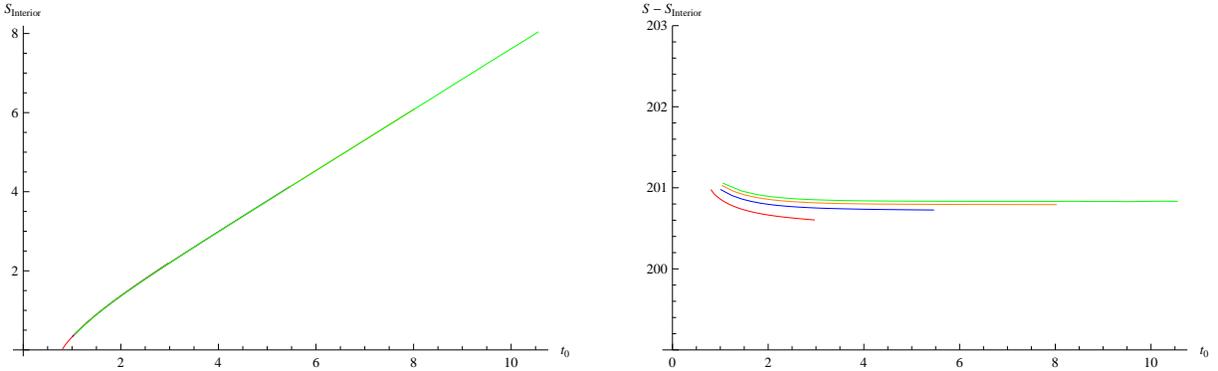}\newline
\caption{$S_{interior}$ (left) and $S-S_{interior}$ (right) for Lifshitz
gravity as functions of $l$ and $t_{0}$. The boundary separations were taken
to be $l=4,6,8,10$ (red, blue, orange, green). Note that for different $l$
their $S_{interior}$ grows almost along the same curve.}
\label{LjLLif}
\end{figure}

\begin{figure}[tbp]
\includegraphics[width=1\textwidth]{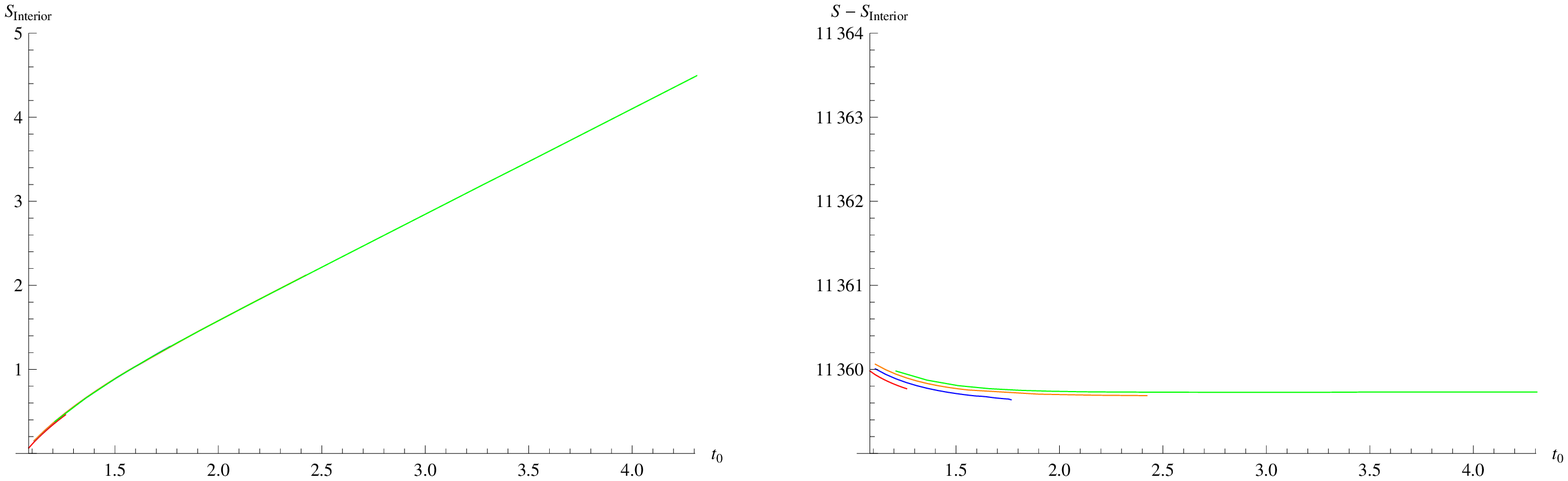}\newline
\caption{$S_{interior}$ (left) and $S-S_{interior}$ (right) for GB gravity
as functions of $l$ and $t_{0}$. The boundary separations were taken to be $%
l=1.5,2,3,7$ (red, blue, orange, green).}
\label{GBInside}
\end{figure}

For other theories with $d>2$, we will resort to numerical methods. Solving
the coordinates $y_{p}$ and $y_{q}$ of crosspoints $p$ and $q$ from%
\begin{equation*}
r_{A}[v(y)]=\frac{1}{z(y)}
\end{equation*}%
and integrating Eq. (\ref{HEE1}) and Eq. (\ref{HEE2}) in the region within $%
y\in \left[ y_{p},y_{q}\right] $, one can obtain the contribution of the HEE
from the extremal surface inside the apparent horizon, see Fig. \ref{LjLd3},
Fig. \ref{LjLLif} and Fig. \ref{GBInside}.

It is clear that the linear growth of HEE in these holographic theories all
comes from the extension of the extremal surfaces in the interior of
apparent horizons.

\section{Conclusion and discussion}

In this paper, using the gauge/gravity duality and Vaidya models, we
investigated the thermalization process in strongly coupling field theories
following a fast and homogeneous energy injection. We detected the
holographic thermalization\ in terms of the HEE and focused on its behavior
with the sufficiently large boundary separation. We studied various
holographic theories, including the d=2, d=3 CFTs, the non-relativistic
scale-invariant theory, and the dual theory of Gauss-Bonnet gravity. We
obtained three \emph{universal} results.

First, for large spatial scale $l$, the evolution of HEE includes an
intermediate stage during which it grows linearly and the growth rate
approaches a $l$-independent constant when the scale increases. Second, the
time-dependent HMI captures the behavior of the linear growth with the $l$%
-independent rate exactly. Third, the linear growth of HEE at large $l$ is
related to the interior of the apparent horizon along the extremal surface.
In particular, all the results are obtained analytically for d=2.

Besides the three main results, we would like to point out an interesting
phenomenon that the growth rate of HEE in the Lifshitz theory initially
decreases with respect to $t_{0}$, which is different with all the other
theories.

We also found that the interior of the event horizon can capture the linear
growth of HEE, see Appendix C. However, it was observed that the extremal
surface inside the event horizon involves the extra part in the vacuum,
which has nothing to do with the behavior of the linear growth. Thus, we
argue that the apparent horizon seems to capture the linear growth behavior
of HEE more exactly than the event horizon does.

At last, we note that the time-dependent HEE can be schematically decomposed
as%
\begin{equation*}
S=S_{EH}+S_{BB}=S_{AH}+S_{VAC}+S_{BB}
\end{equation*}%
where $S_{AH}$, $S_{EH}$, $S_{BB}$ and $S_{VAC}$ denote the partial HEE
contributed by the extremal surface inside apparent horizons, event
horizons, black branes and vacuum, respectively. Such kind of decomposition
suggests that some observables on the boundary have a natural structure
determined by the nontrivial locations in the bulk. It is interesting to
study in the future whether the identification of the structure could be
significant in the dual field theories.

Note added: After we finished this paper, we noticed a recent preprint \cite%
{Liu1305}, in which Liu and Suh have obtained analytically the universal
scaling of the HEE in Vaidya models by studying the geometry around and
inside the event horizon.

\begin{acknowledgments}
SFW would like to thank Hong Liu for very helpful discussion on the local
nature of entanglement propagation. SFW and YQW were supported by National
Natural Science Foundation of China (No. 11275120 and No. 11005054).
\end{acknowledgments}

\appendix

\section{The Vaidya metric for Lifshitz gravity}

The 4-dim infalling shell geometry in asymptotically Lifshitz background
described by the Vaidya metric in Poincar\`{e} coordinates will be given in
this section. The authors of Ref. \cite{Lif} have obtained a static black
hole solution in four dimensions that asymptotically approaches the Lifshitz
spacetime in a system with a strongly-coupled scalar. The action is%
\begin{equation*}
S=\frac{1}{2}\int d^{4}x\sqrt{-g}(R-2\Lambda )-\int d^{4}x\sqrt{-g}\left(
\frac{e^{-2\varphi }}{4}F^{2}+\frac{m^{2}}{2}A^{2}+e^{-2\varphi }-1\right)
\end{equation*}%
where $\Lambda =-\frac{n^{2}+(d-2)n+(d-1)^{2}}{2}$ is the cosmological
constant and $m^{2}=(d-1)n$. The solution of this system is:
\begin{equation}
ds^{2}=-f\frac{dt^{2}}{z^{2n}}+\frac{d\vec{x}^{2}}{z^{2}}+\frac{dz^{2}}{%
fz^{2}},~~A=\frac{f}{\sqrt{2}z^{2}}dt,~~\varphi =-\frac{1}{2}\log \left(
1+z^{2}/z_{H}^{2}\right) ,  \label{Ab}
\end{equation}%
with $f=1-Mz^{2}$ and $n=2$. Using the coordinate transformation
\begin{equation*}
dt=dv+\frac{z^{n-1}}{f}dz,
\end{equation*}%
the metric in Eq. (\ref{Ab}) will become
\begin{equation*}
ds^{2}=-z^{2n}(1-Mz^{2})dv^{2}-2z^{-1-n}dzdv+z^{-2}d\vec{x}^{2}.
\end{equation*}%
Consider Einstein's and Maxwell's equations%
\begin{equation}
T_{\mu \nu }=R_{\mu \nu }-\frac{1}{2}Rg_{\mu \nu }+\Lambda g_{\mu \nu
}-e^{-2\varphi }F_{\mu }^{\gamma }F_{\nu \gamma }-m^{2}A_{\mu }A_{\nu
}+g_{\mu \nu }\left( \frac{e^{-2\varphi }}{4}F^{2}+\frac{m^{2}}{2}%
A^{2}+e^{-2\varphi }-1\right)  \label{Ein}
\end{equation}%
\begin{equation}
J^{\nu }=e^{-2\varphi }\nabla _{\mu }F^{\mu \nu }-m^{2}A^{\nu }.  \label{Max}
\end{equation}%
We find that there exists a Vaidya solution with the form
\begin{eqnarray*}
ds^{2} &=&-z^{2n}\left[ 1-m(v)z^{2}\right] dv^{2}-2z^{-1-n}dzdv+z^{-2}d\vec{x%
}^{2}, \\
A_{\mu } &=&\frac{1-m(v)z^{2}}{\sqrt{2}z^{2}}\delta _{\mu }^{v}+\frac{z^{n-1}%
}{\sqrt{2}z^{2}}\delta _{\mu }^{r}, \\
\varphi (v,z) &=&-\frac{1}{2}\log [1+m(v)z^{2}],
\end{eqnarray*}%
that means, the only nonvanishing components of Eq. (\ref{Ein}) and Eq. (\ref%
{Max}) are given by%
\begin{equation*}
T_{vv}={m}^{\prime }{(v),\;J}_{v}=\sqrt{2}z^{2}{m}^{\prime }{(v).}
\end{equation*}%
One can see that the only difference between the dynamic solution and the
static one is to replace the mass parameter $M$ with a mass function $m(v)$.

\section{Decomposition of static HEE and HMI by numerical methods}

In Ref. \cite{Kundu12122}, it is interesting to see that the static HEE
(both for $d$-dim relativistic CFTs and non-relativistic scale-invariant
theories) at high temperature (i.e. $l\gg \beta $) can be analytically
decomposed as $S=S_{div}+S_{thermal}+S_{finite}+S_{corr}$. Moreover, for $%
l\gg \beta $ and $x\ll \beta $, HMI can be decomposed as $%
I=I_{div}+S_{finite}+I_{corr}$ \cite{Kundu12124}. Now we would like to study
whether there is a similar decomposition in the field theory dual to GB
gravity. However, it seems difficult to analytically calculate HEE in the GB
background since its prescription is nontrivially corrected, see Eq. (\ref%
{HEEGB}). Fortunately, we can achieve the decomposition by numerical fitting.

At the beginning, let us review the analytical decomposition for $d$-dim
CFTs and illustrate the effectiveness of our numerical methods in d=4 CFTs.
It has been found that Eq. (\ref{HEE3}) with $l\gg \beta $ can be decomposed
analytically as

\begin{equation}
S\simeq S_{div}+\frac{\left( r_{H}R\right) ^{d-2}}{4G_{N}^{d+1}}\left(
kr_{H}l+\mathcal{S}_{high}-\mathcal{E}_{1}e^{-\mathcal{E}_{0}r_{H}l}\right) .
\label{Shigh}
\end{equation}%
The divergent term can be gotten by computing Eq. (\ref{HEE3}) in the pure $%
AdS$ spacetime where $M=0$, which gives rise to $S_{div}=\frac{R^{d-2}}{%
4G_{N}^{d+1}}\frac{2}{(d-2)z_{0}^{d-2}}$. The constant $k=1$ and $\mathcal{E}%
_{0}=\sqrt{d(d-1)/2}$. The constants $\mathcal{S}_{high}$ and $\mathcal{E}%
_{1}$ are made of many gamma functions \cite{Kundu12122}. For our aim, we
calculate them for d=4, which are%
\begin{equation}
\mathcal{S}_{high}=-0.665925,\;\mathcal{E}_{1}=1.437285.  \label{ShighEent}
\end{equation}%
Moreover, the HEE (\ref{HEE3}) at low temperature (i.e. $l\ll \beta $) can
be analytically expanded as%
\begin{equation}
S\simeq S_{div}+\frac{R^{d-2}}{4G_{N}^{d+1}}\frac{\mathcal{S}_{0}}{l^{d-2}}%
\left[ 1+\mathcal{S}_{1}\left( r_{H}l\right) ^{d}\right] .  \label{Slow}
\end{equation}%
When d=4,%
\begin{equation}
\mathcal{S}_{0}=-0.320664,\;\mathcal{S}_{1}=-1.763956.  \label{S0S1}
\end{equation}%
Combing Eq. (\ref{Shigh}) and Eq. (\ref{Slow}), one can decompose the HMI to%
\begin{equation*}
I\simeq \frac{\left( r_{H}R\right) ^{d-2}}{4G_{N}^{d+1}}\left[ \frac{-%
\mathcal{S}_{0}}{\left( r_{H}x\right) ^{d-2}}+\mathcal{S}_{high}-kr_{H}x-%
\mathcal{S}_{0}\mathcal{S}_{1}\left( r_{H}x\right) ^{2}\right] ,
\end{equation*}%
when $l\gg \beta $ and $x\ll \beta $.

\begin{figure}[tbp]
\includegraphics[width=1\textwidth]{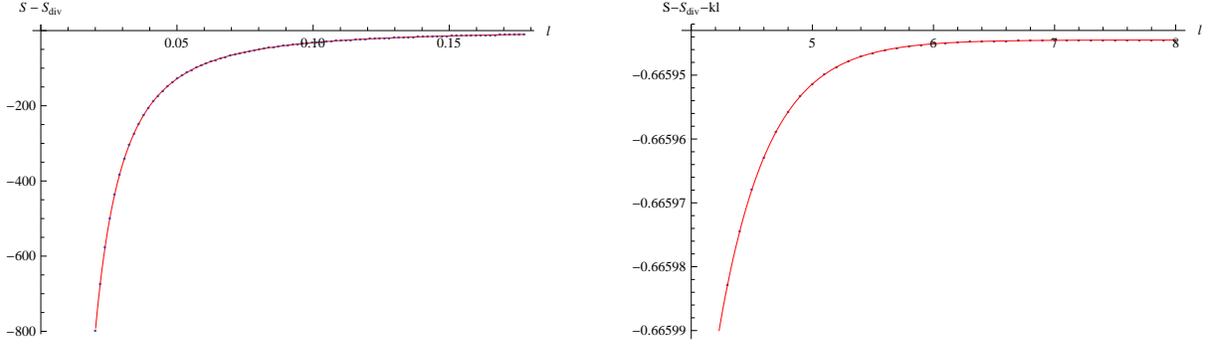}\newline
\caption{Fitting the HEE for d=4 CFTs at small (left) and large (right) $l$.
The blue points are the data of HEE and the red lines are our fitting
functions.}
\label{GRfj}
\end{figure}

Now we invoke the numerical methods. By numerically fitting Eq. (\ref{Shigh}%
) and Eq. (\ref{Slow}) with Eq. (\ref{HEE3}) in the region with large and
small $l$, respectively, we extract the constants%
\begin{eqnarray*}
\mathcal{S}_{high} &=&-0.665944,\;\mathcal{E}_{1}=1.439163. \\
\mathcal{S}_{0} &=&-0.320664,\;\mathcal{S}_{1}=-1.763732,
\end{eqnarray*}%
which match Eq. (\ref{ShighEent}) and Eq. (\ref{S0S1}) very well, see Fig. %
\ref{GRfj}.

\begin{figure}[tbp]
\includegraphics[width=1\textwidth]{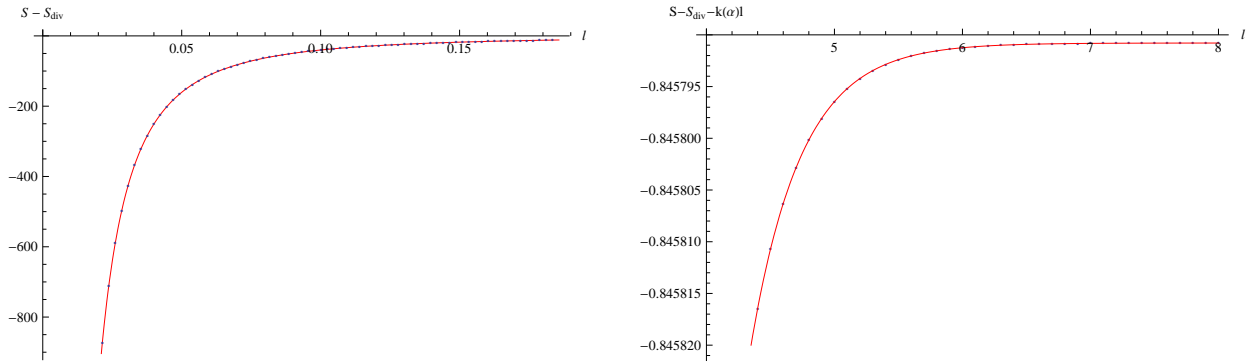}\newline
\caption{Fitting the HEE for the GB gravity at small (left) and large
(right) $l$. The blue points are the data of HEE and the red lines are our
fitting functions.}
\label{GBfj}
\end{figure}

We go further to study the GB gravity. The divergent term of the static HEE
can be calculated as $S_{div}=\frac{R^{2}}{4G_{N}^{5}}\frac{2\alpha
+L_{c}^{2}}{z_{0}^{2}L_{c}^{3}}$. Fitting the HEE with Eq. (\ref{Shigh}) and
Eq. (\ref{Slow}) in different region of $l$, see Fig. \ref{GBfj}, we find
that the constants $k$ and $\mathcal{E}_{0}$ should be equal to the rescaled
values%
\begin{equation*}
k(\alpha )=-\frac{1}{L_{c}^{3}},~~\mathcal{E}_{0}(\alpha )=-\sqrt{\frac{%
d(d-1)}{2}}\frac{1}{L_{c}}.
\end{equation*}%
The remained constants can be obtained for different $\alpha $. For
instance, when $\alpha =0.05$,
\begin{equation*}
\mathcal{S}_{high}(\alpha )=-0.845791,\;\mathcal{S}_{0}(\alpha )=-0.401275,\;%
\mathcal{S}_{1}(\alpha )=-1.276055.
\end{equation*}

Hereto, we have shown that the decomposition of static HEE at low and high
temperature has the general form for $d$-dim CFTs and the field theory dual
to GB gravity. Consequently, the decomposition of HMI is general too.

\section{Linear growth of HEE from the interior of event horizons}

\begin{figure}[tbp]
\centering\includegraphics[width=0.7\textwidth]{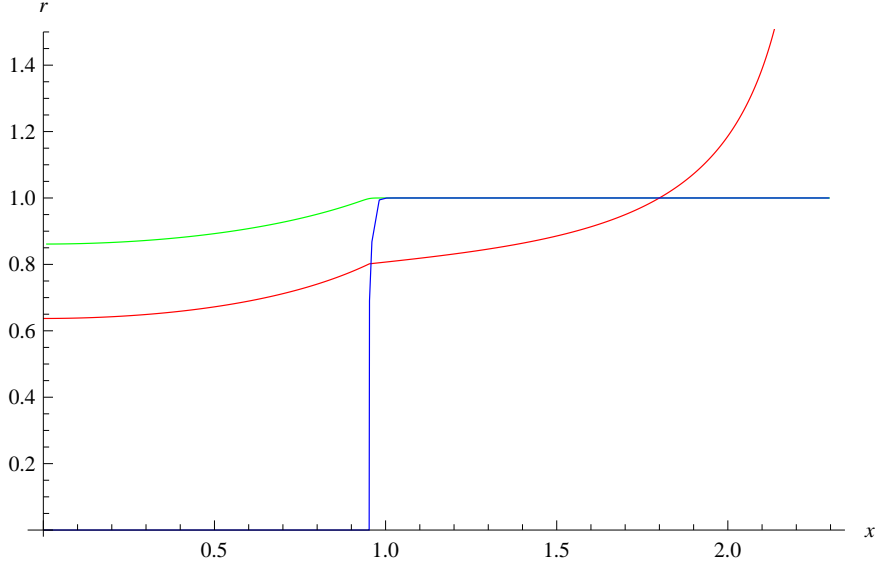}
\caption{The geodesic line (red), the apparent horizon (blue) and the event
horizon (green) as a function of $x$ for fixed $t_{0}=2$ and $l=4.6$ in d=2
CFTs.}
\label{EHCFT2}
\end{figure}

Here we would like to study whether the event horizon can also capture the
behavior of the linear growth of HEE. The event horizon is a null
hypersurface generated by outgoing null geodesics. According to the general
metric (\ref{Vaidya}), its location can be determined by a differential
equation%
\begin{equation*}
f_{1}\left[ z(v),v\right] +2z(v)^{2n-2}f_{2}\left[ z\left( v\right) \right]
\frac{dz(v)}{dv}=0
\end{equation*}%
with the boundary condition of connecting the apparent horizon in the future
of $v=0$ \cite{Poisson04}. By numerical computations, we find that when the
extremal surface intersects the apparent horizon, the event horizon always
lie outside the extremal surface, see Fig. \ref{EHCFT2} for instance.
Thus, in such cases, one can evaluate the part of the HEE contributed by the
extremal surface inside the event horizon by subtracting the part in the
pure black brane from the whole HEE. In the following, we will show that the
subtracted part is independent with $t_{0}$ in the region with large $l$ and
intermediate $t_{0}$. Since we have demonstrated that the contribution of
the extremal surface involving both the vacuum and the pure black brane are
independent with $t_{0}$, we can conclude that the part of HEE contributed
by the extremal surface inside the event horizon grows linearly as the
behavior of complete HEE.

Let us consider d=2 CFTs that can be described analytically. The length of
the desired geodesic outside the event horizon (i.e. the geodesic in the
pure black brane) is%
\begin{eqnarray*}
L_{outside}(l,t_{0}) &=&2\left[ \lambda _{+}^{out}(r_{0})-\lambda
_{+}^{out}(r_{H})\right] \\
&=&2\ln \left( \frac{2r_{0}}{r_{H}}\right) -\ln \left( 1-\frac{r_{s}^{2}}{%
r_{H}^{2}}+\frac{r_{H}\sqrt{r_{sw}^{2}-r_{s}^{2}}}{r_{sw}^{2}}+\frac{%
r_{H}^{2}\left( r_{sw}^{2}-r_{s}^{2}\right) }{4r_{sw}^{4}}\right) .
\end{eqnarray*}%
When $s$ is small, the length can be expanded as%
\begin{equation*}
L_{outside}(l,t_{0})=2\ln \left( \frac{r_{0}}{r_{H}}\right) +2\ln \left(
1+e^{r_{H}t_{0}}\right) -2r_{H}t_{0}+\mathcal{O}(s)^{2}.
\end{equation*}%
Obviously, it approaches a constant $2\log \left( r_{0}/r_{H}\right) $ when $%
t_{0}$ increases.

\begin{figure}[tbp]
\centering\includegraphics[width=1\textwidth]{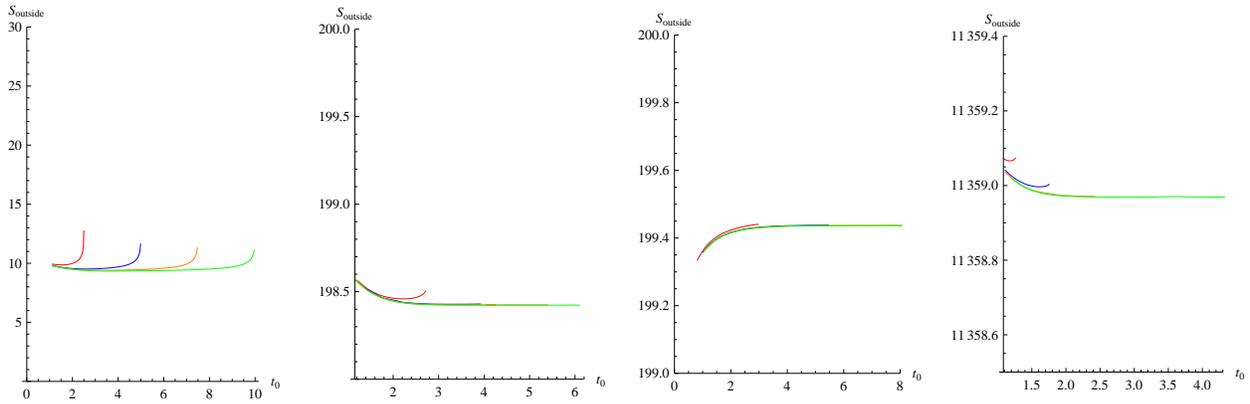}
\caption{The constant contribution to the HEE from the extremal surfaces in
the pure black branes for various holographic theories. From left to right,
they are d=2 CFTs ($l=5,10,15,20$), d=3 CFTs ($l=2,3,4,5$), Lifshitz gravity
($l=4,6,8,10$) and GB gravity ($l=1.5,2,3,7$). In each panel the red, blue,
orange and green lines represent different $l$ from small to large.}
\label{Outside}
\end{figure}
We numerically plot Fig. \ref{Outside} that reveals the constant
contribution to the HEE from the extremal surface in the pure black branes
for various holographic theories.


\begin{thebibliography}{99}
\bibitem{Maldacena9711200} J. M. Maldacena, \emph{The Large N Limit of
Superconformal Field Theories and Supergravity}, \emph{Adv. Theor. Math.
Phys.} \textbf{2} (1998) 231 [hep-th/9711200]; S. S. Gubser, I. R. Klebanov,
A. M. Polyakov, \emph{Gauge Theory Correlators from Non-Critical String
Theory}, \emph{Phys. Lett. }\textbf{B 428 }(1998) 105 [hep-th/9802109]; E.
Witten, \emph{Anti De Sitter Space And Holography}, \emph{Adv. Theor. Math.
Phys.} \textbf{2} (1998) 253 [hep-th/9802150].

\bibitem{MullerReview} See following reviews: F. Gelis, \emph{The early
stages of a high energy heavy ion collision, J. Phys. Conf. Ser.} 381 (2012)
012021 [arXiv:1110.1544]; B. M\"{u}ller and A. Sch\"{a}fer, \emph{Entropy
creation in relativistic heavy ion collisions}, arXiv:1110.2378; E. Iancu,
\emph{QCD in heavy ion collisions}, arXiv:1205.0579.

\bibitem{Son01} R. Baier, A. H. Mueller, D. Schi and D. T. Son, \emph{%
Bottom-up thermalization in heavy ion collisions}, \emph{Phys. Lett.}
\textbf{B 502} (2001) 51 [hep-ph/0009237]; A. H. Mueller, A. I. Shoshi and
S. M. H. Wong, \emph{A Possible Modified "bottom-up" Thermalization in Heavy
Ion Collisions}, \emph{Phys. Lett.} \textbf{B 632} (2006) 257
[hep-ph/0505164].

\bibitem{Polkovnikov} See following reviews: S. Mondal, D. Sen and K.
Sengupta, \emph{Non-equilibrium dynamics of quantum systems: order parameter
evolution, defect generation, and qubit transfer}, arXiv:0908.2922; J.
Dziarmaga, \emph{Dynamics of a quantum phase transition and relaxation to a
steady state}, arXiv:0912.4034; A. Polkovnikov, K. Sengupta, A. Silva and M.
Vengalattore, \emph{Nonequilibrium dynamics of closed interacting quantum
systems}, \emph{Rev. Mod. Phys.} \textbf{83} (2011) 863 [arXiv:1007.5331].

\bibitem{Nielsen00} M. A. Nielsen and I. L. Chuang, \emph{Quantum
computation and quantum information,} Cambridge University Press, 2000.

\bibitem{Wen0510} P. Calabrese and J. L. Cardy, \emph{Entanglement entropy
and quantum field theory, J. Stat. Mech.} \textbf{0406} (2004) P06002
[hep-th/0405152]; A. Kitaev and J. Preskill, \emph{Topological entanglement
entropy, Phys. Rev. Lett.} \textbf{96} (2006) 110404 [hep-th/0510092]; M.
Levin and X. G. Wen, \emph{Detecting Topological Order in a Ground State
Wave Function}, \emph{Phys. Rev. Lett.} \textbf{96} (2006) 110405
[cond-mat/0510613]; B. Hsu, M. Mulligan, E. Fradkin and E. A. Kim, \emph{%
Universal entanglement entropy in 2D conformal quantum critical points},
\emph{Phys. Rev.} \textbf{B 79} (2009) 115421 [arXiv:0812.0203].

\bibitem{HEE} S. Ryu and T. Takayanagi, \emph{Holographic Derivation of
Entanglement Entropy from AdS/CFT, Phys. Rev. Lett.} \textbf{96} (2006)
181602 [hep-th/0603001]; S. Ryu and T. Takayanagi, \emph{Aspects of
holographics entanglement entropy, JHEP} \textbf{08} (2006) 045
[hep-th/0605073].

\bibitem{HEErev} See a recent review: T. Takayanagi, \emph{\ Entanglement
Entropy from a Holographic Viewpoint, Class. Quant. Grav.} \textbf{29}
(2012) 153001 [arXiv:1204.2450].

\bibitem{Maldacena1304} A. Lewkowycz and J. Maldacena, \emph{Generalized
gravitational entropy}, arXiv:1304.4926.

\bibitem{Raamsdonk09} M. Van Raamsdonk, \emph{Comments on quantum gravity
and entanglement}, arXiv:0907.2939; M. Van Raamsdonk, \emph{Building up
spacetime with quantum entanglement}, \emph{Gen. Rel. Grav.} \textbf{42}
(2010) 2323 [arXiv:1005.3035].

\bibitem{Myers1102} H. Casini, M. Huerta and R. C. Myers, \emph{Towards a
derivation of holographic entanglement entropy}, \emph{JHEP} \textbf{05}
(2011) 036 [arXiv:1102.0440]; E. Bianchi and R. C. Myers, \emph{On the
Architecture of Spacetime Geometry}, arXiv:1212.5183.

\bibitem{Hubeny1204} V. E. Hubeny and M. Rangamani, \emph{Causal Holographic
Information}, arXiv:1204.1698;

\bibitem{Hubeny1302} V. E. Hubeny, M. Rangamani, E. Tonni, \emph{%
Thermalization of Causal Holographic Information}, arXiv:1302.0853.

\bibitem{Swingle09} B. Swingle, \emph{Entanglement Renormalization and
Holography}, \emph{Phys. Rev.} \textbf{D 86} (2012) 065007
[arXiv:0905.1317]; \emph{Constructing holographic spacetimes using
entanglement renormalization}, arXiv:1209.3304.

\bibitem{Vilaplana11} J. Molina-Vilaplana and P. Sodano, \emph{Holographic
View on Quantum Correlations and Mutual Information between Disjoint Blocks
of a Quantum Critical System}, \emph{JHEP} \textbf{10} (2011) 011
[arXiv:1108.1277].

\bibitem{Bala1108} V. Balasubramanian, M. B. McDermott and M. Van Raamsdonk,
\emph{Momentum-space entanglement and renormalization in quantum field theory%
}, arXiv:1108.3568; H. Matsueda, \emph{Scaling of entanglement entropy and
hyperbolic geometry}, arXiv:1112.5566; M. Ishihara, F. L. Lin and B. Ning,
\emph{Refined Holographic Entanglement Entropy for the AdS Solitons and AdS
black Holes}, arXiv:1203.6153; H. Matsueda, M. Ishihara and Y. Hashizume,
\emph{Tensor Network and Black Hole}, arXiv:1208.0206.

\bibitem{Takayanagi12} M. Nozaki, S. Ryu and T. Takayanagi, \emph{%
Holographic Geometry of Entanglement Renormalization in Quantum Field
Theories}, \emph{JHEP} \textbf{10} (2012) 193 [arXiv:1208.3469].

\bibitem{Takayanagi1302} M. Nozakia, T. Numasawa and T. Takayanagi, \emph{%
Holographic Local Quenches and Entanglement Density}, arXiv:1302.5703.

\bibitem{Maldacena13} T. Hartman and J. Maldacena, \emph{Time Evolution of
Entanglement Entropy from Black Hole Interiors}, arXiv:1303.1080.

\bibitem{MERA} G. Vidal, \emph{Entanglement renormalization, Phys. Rev. Lett.%
} \textbf{99} (2007) 220405 [cond-mat/0512165]; G. Vidal, \emph{Entanglement
renormalization: an introduction,} arXiv:0912.1651; J. Haegeman, T. J.
Osborne, H. Verschelde and F. Verstraete, \emph{Entanglement renormalization
for quantum fields, Phys. Rev. Lett.} \textbf{110} (2013) 100402
[arXiv:1102.5524]; G. Evenbly and G. Vidal, \emph{Quantum Criticality with
the Multi-scale Entanglement Renormalization Ansatz,} arXiv:1109.5334.

\bibitem{Hubeny0705} V. E. Hubeny, M. Rangamani and T. Takayanagi, \emph{A
Covariant Holographic Entanglement Entropy Proposal, JHEP} \textbf{07}
(2007) 062 [arXiv:0705.0016].

\bibitem{Wolf07} M. M. Wolf, F. Verstraete, M. B. Hastings, and J. I. Cirac,
\emph{Area laws in quantum systems: Mutual information and correlations,
Phys. Rev. Lett.} \textbf{100} (2008) 070502 [arXiv:0704.3906].

\bibitem{Swingle1010} B. Swingle, \emph{Mutual information and the structure
of entanglement in quantum field theory}, arXiv:1010.4038.

\bibitem{Furukawa08} S. Furukawa, V. Pasquier and J. Shiraishi, \emph{Mutual
Information and Boson Radius in c=1 Critical Systems in One Dimension},
\emph{Phys. Rev. Lett.} \textbf{102} (2009) 170602 [arXiv:0809.5113]; P.
Calabrese, J. Cardy and E. Tonni, \emph{Entanglement entropy of two disjoint
intervals in conformal field theory}, \emph{J. Stat. Mech.} (2009) P11001
[arXiv:0905.2069 [hep-th]]; P. Calabrese, J. Cardy and E. Tonni, \emph{%
Entanglement entropy of two disjoint intervals in conformal field theory II}%
, \emph{J. Stat. Mech.} \textbf{1101} (2011) P01021 [arXiv:1011.5482].

\bibitem{Headrick06} M. Headrick,\emph{\ Entanglement Renyi entropies in
holographic theories, Phys. Rev.} \textbf{D 82} (2010) 126010
[arXiv:1006.0047].

\bibitem{Hubeny0711} V. E. Hubeny and M. Rangamani, \emph{Holographic
entanglement entropy for disconnected regions}, \emph{JHEP} \textbf{03}
(2008) 006 [arXiv:0711.4118].

\bibitem{Tonni1011} E. Tonni, \emph{Holographic entanglement entropy: near
horizon geometry and disconnected regions}, \emph{JHEP} \textbf{05} (2011)
004 [arXiv:1011.0166].

\bibitem{BalaMI} V. Balasubramanian, A. Bernamonti, N. Copland, B. Craps and
F. Galli, \emph{Thermalization of mutual and tripartite information in
strongly coupled two dimensional conformal field theories}, \emph{Phys. Rev.}
\textbf{D 84} (2011) 105017 [arXiv:1110.0488].

\bibitem{Tonni11} A. Allais and E. Tonni, \emph{Holographic evolution of the
mutual information}, \emph{JHEP} \textbf{01} (2012) 102 [arXiv:1110.1607].

\bibitem{Callan12} R. Callan, J. He, and M. Headrick, \emph{Strong
subadditivity and the covariant holographic entanglement entropy formula},
\emph{JHEP} \textbf{06} (2012) 081 [arXiv:1204.2309].

\bibitem{Kundu12122} W. Fischler and S. Kundu, \emph{Strongly Coupled Gauge
Theories: High and Low Temperature Behavior of Non-local Observables},
arXiv:1212.2643.

\bibitem{Kundu12124} W. Fischler, A. Kundu and S. Kundu, \emph{Holographic
Mutual Information at Finite Temperature}, arXiv:1212.4764.

\bibitem{Bhattacharyya0904} S. Bhattacharyya and S. Minwalla, \emph{\ Weak
Field Black Hole Formation in Asymptotically AdS Spacetimes}, \emph{JHEP}
\textbf{09} (2009) 034 [arXiv:0904.0464].

\bibitem{quench} S. R. Das, T. Nishioka and T. Takayanagi, \emph{Probe
Branes, Time-dependent Couplings and Thermalization in AdS/CFT}, \emph{JHEP}
\textbf{07} (2010) 071 [arXiv:1005.3348]; H. Ebrahim and M. Headrick, \emph{%
Instantaneous Thermalization in Holographic Plasmas}, BRX-TH 624
[arXiv:1010.5443]; D. Garfinkle and L. A. Pando Zayas, \emph{Rapid
Thermalization in Field Theory from Gravitational Collapse}, \emph{Phys. Rev.%
} \textbf{D 84} (2011) 066006 [arXiv:1106.2339]; S. R. Das, \emph{%
Holographic Quantum Quench}, \emph{J. Phys. Conf. Ser.} \textbf{343} (2012)
012027 [arXiv:1111.7275]; A. Buchel, L. Lehner and R. C. Myers, \emph{%
Thermal quenches in $\mathcal{N}=2^{*}$ plasmas}, \emph{JHEP} \textbf{08}
(2012) 049 [arXiv:1206.6785]; X. Gao, A. M. Garcia-Garcia, H. B. Zeng and H.
Q. Zhang, \emph{Lack of thermalization in holographic superconductivity},
arXiv:1212.1049; W. H. Baron, D. Galante and M. Schvellinger, \emph{Dynamics
of holographic thermalization}, \emph{JHEP} \textbf{03} (2013) 070
[arXiv:1212.5234].

\bibitem{Arrastia10} J. Abajo-Arrastia, J. Aparicio and E. Lopez, \emph{%
Holographic Evolution of Entanglement Entropy}, \emph{JHEP} \textbf{11}
(2010) 149 [arXiv:1006.4090].

\bibitem{Aparicio11} J. Aparicio and E. Lopez, \emph{Evolution of Two-Point
Functions from Holography}, \emph{JHEP} \textbf{12} (2011) 082
[arXiv:1109.3571].

\bibitem{Johnson11} T. Albash and C. V. Johnson, \emph{Evolution of
Holographic Entanglement Entropy after Thermal and Electromagnetic Quenches}%
, \emph{New J. Phys.} \textbf{13} (2011) 045017 [arXiv:1008.3027].

\bibitem{Bala11} V. Balasubramanian, A. Bernamonti, J. de Boer, N. Copland,
B. Craps, E. Keski-Vakkuri, B. M\"{u}ller and A. Sch\"{a}fer, \emph{%
Thermalization of Strongly Coupled Field Theories}, \emph{Phys. Rev. Lett.}
\textbf{106} (2011) 191601 [arXiv:1012.4753]; V. Balasubramanian, A.
Bernamonti, J. de Boer, N. Copland, B. Craps, E. Keski-Vakkuri, B. M\"{u}%
ller and A. Sch\"{a}fer, \emph{Holographic Thermalization}, \emph{Phys. Rev.}
\textbf{D 84} (2011) 026010 [arXiv:1103.2683].

\bibitem{Keranen11} V. Keranen, E. Keski-Vakkuri, L. Thorlacius, \emph{%
Thermalization and entanglement following a non-relativistic holographic
quench}, \emph{Phys. Rev.} \textbf{D 85} (2012) 026005 [arXiv:1110.5035].

\bibitem{Galante1205} D. Galante and M. Schvellinger, \emph{Thermalization
with a chemical potential from AdS spaces}, \emph{JHEP} \textbf{07} (2012)
096 [arXiv:1205.1548]; E. Caceres and A. Kundu, \emph{Holographic
Thermalization with Chemical Potential}, \emph{JHEP} \textbf{09} (2012) 055
[arXiv:1205.2354]; E. Caceres, A. Kundu, J. F. Pedrazab, W. Tangarife, \emph{%
Strong Subadditivity, Null Energy Condition and Charged Black Holes},
arXiv:1304.3398.

\bibitem{Liuwenbiao} X. X. Zeng and W. B. Liu, \emph{Holographic
thermalization in Gauss-Bonnet gravity}, arXiv:1305.4841.

\bibitem{Lin08} S. Lin and E. Shuryak, \emph{Toward the AdS/CFT Gravity Dual
for High Energy Collisions.3. Gravitationally Collapsing Shell and
Quasiequilibrium}, \emph{Phys. Rev.} \textbf{D 78} (2008) 125018
[arXiv:0808.0910].

\bibitem{Cardy05} P. Calabrese and J. Cardy, \emph{Evolution of Entanglement
Entropy in One-dim Systems,\ J. Stat. Mech.} \textbf{0504} (2005) P04010
[cond-mat/0503393].

\bibitem{Cardy07} P. Calabrese and J. Cardy, \emph{Quantum Quenches in
Extended Systems, J. Stat. Mech.} \textbf{0706} (2007) P06008
[arXiv:0704.1880].

\bibitem{Latora99} P. Billingsley, \emph{Ergodic Theory and Information},
Wiley, New York, 1965; V. Latora and M. Baranger, \emph{Kolmogorov-Sinai
Entropy Rate versus Physical Entropy}, \emph{Phys. Rev. Lett} \textbf{82}
(1999) 520 [chao-dyn/9806006].

\bibitem{Muller00} J. Bolte, B. Mu\"{u}ller, and A. Sch\"{a}fer, \emph{%
Ergodic properties of classical SU(2) lattice gauge theory}, \emph{Phys. Rev.%
} \textbf{D 61} (2000) 054506 [hep-lat/9906037]; T. Kunihiro et al., \emph{%
Chaotic behavior in classical Yang-Mills dynamics}, \emph{Phys. Rev.}
\textbf{D 82} (2010) 114015 [arXiv:1008.1156].

\bibitem{Fursaev06} D. V. Fursaev, \emph{Proof of the holographic formula
for entanglement entropy}, JHEP \textbf{0609} (2006) 018
[arXiv:hep-th/0606184].

\bibitem{GB1} J. de Boer, M. Kulaxizi, A. Parnachev, \emph{Holographic
Entanglement Entropy in Lovelock Gravities}, \emph{JHEP} \textbf{04} (2011)
025 [arXiv:1101.5781].

\bibitem{Myers1101} L. Y. Hung, R. C. Myers and M. Smolkin, \emph{On
Holographic Entanglement Entropy and Higher Curvature Gravity}, \emph{JHEP}
\textbf{04} (2011) 025 [arXiv:1101.5813].

\bibitem{Sinha1305} See the recent proof of EE proposal in GB gravity: A.
Bhattacharyya, A. Kaviraj and A. Sinha, \emph{Entanglement entropy in higher
derivative holography}, arXiv:1305.6694.

\bibitem{Asplund11} C. T. Asplund and S. G. Avery, \emph{Evolution of
Entanglement Entropy in the D1-D5 Brane System}, \emph{Phys. Rev.} \textbf{D
84} (2011) 124053 [arXiv:1108.2510].

\bibitem{Basu12} P. Basu and S. R. Das, \emph{Quantum Quench across a
Holographic Critical Point}, \emph{JHEP} \textbf{01} (2012) 103
[arXiv:1109.3909].

\bibitem{Buchel13} A. Buchel, L. Lehner, R. C. Myers and A. van Niekerk,
\emph{Quantum quenches of holographic plasmas}, arXiv:1302.2924.

\bibitem{Kobayashi05} T. Kobayashi, \emph{A Vaidya-type radiating solution
in Einstein-Gauss-Bonnet gravity and its application to braneworld}, \emph{%
Gen. Rel. Grav.} \textbf{37} (2005) 1869 [gr-qc/0504027]; H. Maeda, \emph{%
Effects of Gauss-Bonnet term on the final fate of gravitational collapse},
\emph{Class. Quant. Grav.} \textbf{23} (2006) 2155 [gr-qc/0504028]; A. E.
Dominguez and E. Gallo, \emph{Radiating black hole solutions in
Einstein-Gauss-Bonnet gravity}, \emph{Phys. Rev.} \textbf{D 73} (2006)
064018 [gr-qc/0512150].

\bibitem{Lif} K. Balasubramanian and J. McGreevy, \emph{An analytic Lifshitz
black hole}, \emph{Phys. Rev.} \textbf{D 80} (2009) 104039 [arXiv:0909.0263].

\bibitem{Solodukhin} S. N. Solodukhin, \emph{Entanglement Entropy in
Non-Relativistic Field Theories}, \emph{JHEP} \textbf{04} (2010) 101
[arXiv:0909.0277]; D. Nesterov and S. N. Solodukhin, \emph{Gravitational
effective action and entanglement entropy in UV modified theories with and
without Lorentz symmetry}, \emph{Nucl. Phys.} \textbf{B 842} (2011) 141
[arXiv:1007.1246].

\bibitem{Hawking73} S. W. Hawking and G. F. R. Ellis, \emph{The Large scale
structure of space-time}, Cambridge University Press, Cambridge, 1973.

\bibitem{Chesler0812} P. M. Chesler and L. G. Yaffe, \emph{Horizon formation
and far-from-equilibrium isotropization in supersymmetric Yang-Mills plasma}%
, \emph{Phys. Rev. Lett.} \textbf{102} (2009) 211601 [arXiv:0812.2053]; P.
M. Chesler and L. G. Yaffe, \emph{Boost invariant flow, black hole
formation, and far-from-equilibrium dynamics in N = 4 supersymmetric
Yang-Mills theory}, \emph{Phys. Rev.} \textbf{D 82} (2010) 026006
[arXiv:0906.4426].

\bibitem{Hubeny0902} P. Figueras, V. E. Hubeny, M. Rangamani and S. F. Ross,
\emph{Dynamical black holes and expanding plasmas}, \emph{JHEP} \textbf{04}
(2009) 137 [arXiv:0902.4696].

\bibitem{Takayanagi08} T. Takayanagi and T. Ugajin, \emph{Measuring Black
Hole Formations by Entanglement Entropy via Coarse-Graining}, \emph{JHEP}
\textbf{11} (2010) 054 [arXiv:1008.3439].

\bibitem{Brown86} J. D. Brown and M. Henneaux, \emph{Central Charges In The
Canonical Realization Of Asymptotic Symmetries: an example from three
dimensional gravity}, \emph{Commun. Math. Phys.} 104 (1986) 207.

\bibitem{Poisson04} E. Poisson, \emph{A Relativist's Toolkit}, Cambridge
University Press, 2004.

\bibitem{Liu1305} H. Liu and S. J. Suh, \emph{Entanglement Tsunami:
Universal Scaling in Holographic Thermalization, }arXiv:1305.7244.
\end{thebibliography}
\end{document}